\documentclass[twocolumn,prd,unsortedaddress,superscriptaddress,letterpaper]{revtex4}

\usepackage{graphicx}
\usepackage{dcolumn}
\usepackage{amsmath}
\usepackage{amssymb}
\usepackage{pslatex}
\usepackage{epsfig}

\def\onehalf{ {\textstyle{\frac{1}{2}}}}

\newcommand{\be}{\begin{equation}}
\newcommand{\ee}{\end{equation}}

\def\rBulge{{r_b}}

 \def\calG{{\sl G}}

 \def\calS{{\cal S}}

\def\onehalf{{\textstyle{ \frac{1}{2}}}}

\def\sbar{\bar s}

\def\bfgamma{\mbox{\boldmath$\gamma$}}

\def\bfgamma{\mbox{\boldmath$\Upsilon$}}

\def\reals{\mbox{I$\!$R}}
\def\realR{\mbox{I$\!$R}}
\def\nGores{{n_{g}}}

\newcommand{\bfa}{ {\bf a}}
\newcommand{\bfb}{ {\bf b}}
\newcommand{\bft}{ {\bf t}}
\newcommand{\bfx}{ {\bf x}}

\newcommand{\bfN}{ {\bf N}}

\newcommand{\beq}{\begin{equation}}
\newcommand{\eeq}{\end{equation}}

\begin{document}

\title{The ExaVolt Antenna: A Large-Aperture, Balloon-embedded Antenna \\
for Ultra-high Energy Particle Detection}

\author{P.~W.~Gorham} 
\affiliation{University of Hawaii at Manoa, 
Department of Physics and Astronomy,
Honolulu, Hawaii 96822}
\author{F.~E.~Baginski}
\affiliation{ Department of Mathematics,
 The George Washington University,
 Washington, DC 20052}
\author{P.~Allison} 
\affiliation{University of Hawaii at Manoa, 
Department of Physics and Astronomy,
Honolulu, Hawaii 96822}
\affiliation{Presently at: Department of Physics, Ohio State University
Columbus, OH.}
\author {K.~M.~Liewer}
\affiliation{Jet Propulsion Laboratory, Pasadena, CA}
\author{C.~Miki}
\affiliation{University of Hawaii at Manoa, 
Department of Physics and Astronomy,
Honolulu, Hawaii 96822}
\author{B.~Hill}
\affiliation{University of Hawaii at Manoa, 
Department of Physics and Astronomy,
Honolulu, Hawaii 96822}
\author{G.~S.~Varner}
\affiliation{University of Hawaii at Manoa, 
Department of Physics and Astronomy,
Honolulu, Hawaii 96822}

\begin{abstract}
We describe the scientific motivation, experimental basis, design methodology, and
simulated performance of the ExaVolt Antenna (EVA) mission, and planned ultra-high
energy (UHE) particle observatory under development for NASA's suborbital super-pressure
balloon program in Antarctica. EVA will improve over ANITA's integrated totals -- 
the current state-of-the-art in UHE suborbital
payloads -- by 1-2 orders of magnitude in a single flight. The design is based on 
a novel application of toroidal reflector optics which utilizes a super-pressure
balloon surface, along with a feed-array mounted on an inner membrane, to create
an ultra-large radio antenna system with a synoptic view of the Antarctic ice
sheet below it. Radio impulses arise via the Askaryan effect  when UHE neutrinos interact
within the ice, or via geosynchrotron emission when UHE cosmic rays interact
in the atmosphere above the continent. EVA's instantaneous antenna aperture 
is estimated to be several hundred m$^2$ for detection of these events within
a 150-600~MHz band. For standard cosmogenic UHE neutrino models, EVA should detect
of order 30 events per flight in the EeV energy regime. For UHE cosmic rays,
of order 15,000 geosynchrotron events would be detected in total, several hundred above 10~EeV, and
of order 60 above the GZK cutoff energy.
\end{abstract}

\maketitle

\section{Introduction}
A wide range of efforts are currently aimed at measurement
of the absolute flux levels and energy spectral characteristics of
the ultra-high energy (UHE) cosmogenic neutrino flux
~\cite{Cosmonu,Seckel05,Engel01,Hill85}.
This flux of extragalactic neutrinos in the Exavolt (1 EeV = $10^{18}$~eV) 
energy range must be present at a level that is constrained by the
known existence, emerging composition, and unknown cosmic evolution of the
sources of the ultra-high energy cosmic rays (UHECR).
Recent data from the completed HiRes and ongoing Auger UHECR observatories has 
provided compelling evidence for the first time of the process by which the
UHE cosmogenic neutrinos are generated~\cite{HiRes06,Auger08}, the interaction
of UHECRs with the cosmic microwave background radiation (CMBR), which
produces unstable secondaries decaying into neutrinos, 
as first elucidated in the 1960's by Greisen~\cite{Greisen} and
independently by Zatsepin and Kuzmin~\cite{ZK}, and 
whose resulting signature in the UHECR energy spectrum is now known as the GZK cutoff.

Despite the apparent observation of the GZK cutoff, and the 
corollary that
cosmogenic neutrinos are guaranteed to be present with some certainty, 
the mystery surrounding
the UHECR particles has certainly not diminished. 
If they are accelerated in GRB events~\cite{GRB},
or close to the central engines in cosmically distributed AGN~\cite{AGN} 
we may never directly detect their sources via 
charged-particle astronomy, since
the directions would have to be untangled from the unknown magnetic
fields in the universe, perhaps out to distances of several
hundred Mpc, a daunting task. However, it is in attacking this part
of the problem that neutrino measurements are most critical, since
{\em every GZK neutrino detected must in fact point back to a UHECR
source.} This statement arises as a corollary to the production mechanism
described above; that is, since the daughter neutrino momenta closely
match that of the parent UHECR particles in the lab frame, 
their angle of arrival is nearly identical to the source
direction as observed from earth. 
Although neutrino astronomy is at a very early phase, we can
look forward to the promise of high precision measurements of
such neutrino sources as a completely new branch of astrophysical
observations, independent and complementary to all others.

We present here a concept study for an ultra-sensitive ultra-high energy neutrino
and UHECR observatory, based on detection of radio impulses from either
the Askaryan effect in a neutrino-initiated cascade within a suitable dielectric for neutrinos,
or via geosynchrotron emission from a UHECR-initiated giant air shower in
the Earth's magnetic field. The detector described here, which we denote as the
ExaVolt Antenna (EVA) is under consideration as
a National Aeronautics and Space Administration (NASA) Super-Pressure Balloon
(SPB) mission, and is currently in a technology development phase.
In what follows, we build upon the successful results of the Antarctic Impulsive
Transient Antenna (ANITA)~\cite{ANITA2}, which is the primary precursor to
EVA. ANITA was the first NASA payload to attempt the observation
of cosmogenic neutrinos, and can thus be characterized as a discovery or `pathfinder' mission,
capable of much higher sensitivity than any previous experiment in the 
EeV-ZeV ($10^{18-21}$~eV) energy range of the neutrino spectrum.
EVA builds on ANITA's pioneering approach, extending the
energy threshold an order of magnitude further into the heart
of the expected neutrino energy spectrum. Where ANITA's combined sensitivity
may be adequate to detect at most a handful of BZ neutrino events over
several flights, thus establishing first order parameters for the
flux and spectral energy distribution, our design goal for EVA 
is to increase the sensitivity by as much as two orders of magnitude,
as measured by the number of detected neutrinos, leading to
as many as a hundred or more events per flight. In addition, ANITA's recent
detection of UHECRs via geosynchrotron emission seen in reflection from the
Antarctic ice sheets~\cite{ANITA-UHECR} indicates that a mission with much higher sensitivity will
also record a substantial number of radio-detected UHECR events.

\section{Technical Basis \& Methodology}

To set the stage for a complete understanding of the EVA methodology, 
we first describe the scientific basis for the approach, 
since this has undergone rapid development over the last decade.

\subsection{Theoretical and Experimental Basis for the approach.}
	
\paragraph*{The Askaryan Effect.}
Both ANITA and EVA rely on a property of electromagnetic showers in dielectric
media induced by interactions of 
high energy particles which has become known as the Askaryan effect. 
Particles of primary energy well above the electron-positron 
pair production threshold of order 1~MeV can produce such pairs with 
many generations of daughter particles. Such secondaries themselves 
will create sub-showers, and for primary particles of energies well
above the so-called critical energy (where the pair-production gain
exceeds the competing losses due to ionization of the material),
the shower  grows to a broad
maximum of $e^+e^-\gamma$ with a total number of order the
initial particle energy measured in GeV. However, because of the
asymmetry of $e^+$ interactions in the medium compared to $e^-$, combined
with the available of atomic electrons which can be upscattered into
the shower, a net negative charge excess develops rapidly at the
10-20\% level. 

G. Askaryan \cite{Ask62} first described the process
and its potential use for detection of high energy particles in the
early 1960's, but it was not experimentally confirmed until a series of tests
at the Argonne Wakefield Accelerator (AWA) 
and the Stanford Linear Accelerator (SLAC)
within the last decade~\cite{Gor00,Sal01,Gor05,Gor06}. The effect
has now been measured under a variety of conditions and the theory
has been confirmed to good precision in three different solid dielectric
media, silica sand, salt, and ice.~\cite{Sal01,Gor05,Gor06,Mio06}.
First mention of the idea of radio neutrino searches in Antarctic
ice (via a surface array) is attributable to Gusev \& Zheleznykh~\cite{Gusev}
in 1984.

The development of the negative charge asymmetry means that the
propagating shower, which consists of a compact charged-particle and
photon bunch with dimensions of order a few mm in longitudinal 
thickness and a few cm in transverse width, radiates coherently 
(eg., as if it were a single charge) 
for wavelengths that are larger than the
projected dimensions of the bunch. Given the compact sizes
of such bunches in solid media, the coherence extends well into the
microwave regime~\cite{Mio06}, and the emission forms an
unbroken continuum covering all radio bands below $\sim 10$~GHz.
However, for such radiation to be observable by a detector,
the dielectric target medium must also be transparent over the 
frequency band of the emission. 

The common characteristics of such radiation that are important 
to its use in high energy particle detection by experiments 
such as ANITA and EVA are (1) the extremely broadband spectrum, as described
above; (2) inherent 100\% linear polarization, lying in the
plane defined by the Poynting vector and the shower momentum
vector; (3) the Cherenkov
cone geometry, with a width that is frequency dependent; and
(4) the highly
impulsive nature of the Cherenkov wavefront as received
in the time domain. Each of these characteristics has been
progressively illuminated within the last eight years by the
various experiments at AWA and SLAC. Most recently, 
SLAC testbeam experiment T486, performed in the summer of
2006 as part of the ANITA-I flight calibration, has now for the
first time established the behavior of the Askaryan effect in
ice~\cite{Gor06}.
This data thus provides a validation of the Askaryan effect
in ice, retiring any remaining doubts
or risk that ice might deviate in its response to the Askaryan process.

\subsection{The EVA Design}

\begin{figure}[htb!]
\centering
\includegraphics[width=3.3in]{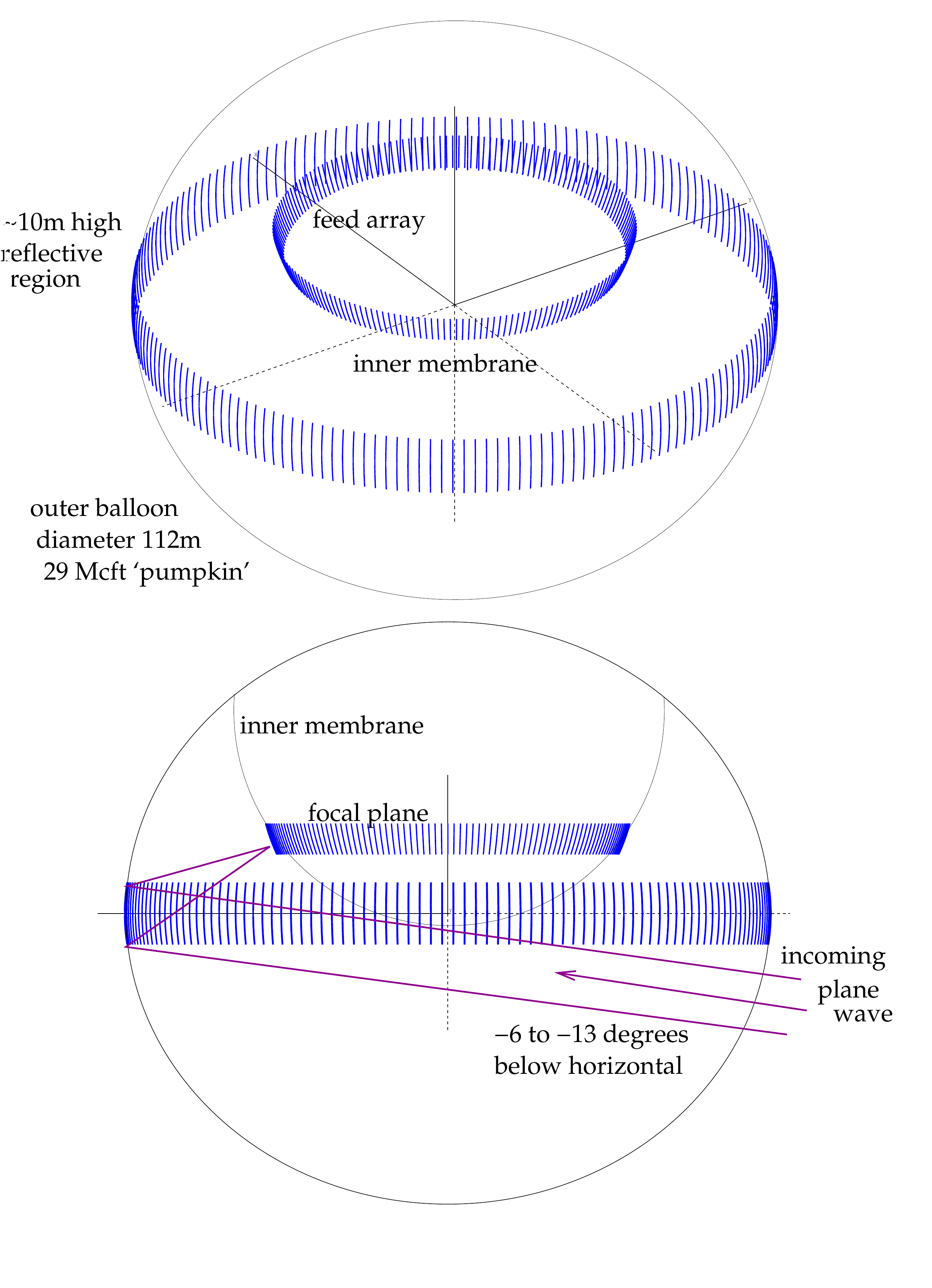}
\begin{small}
\caption{The physical geometry of the reflective outer section and
the inner balloon feed array for the antenna modeling.
Here we have modeled a SPB super-pressure balloon with
about 29M cubic feet of volume and a equatorial
radius of about 56m. The inner membrane has a radius of about
32 m.}
\label{balloon1}
\end{small}
\end{figure}

EVA will deploy the largest-aperture physical telescope ever flown
on a balloon payload, several thousand square meters, to be used
to extend the radio-frequency collection power for neutrino impulses
by a factor of 100 or more over any previous experiment. This will be 
accomplished using a portion of the physical balloon surface itself as the 
radio reflector, with a toroidal geometry as indicated in Fig.\ref{balloon1},
where an approximately 10 m high section of the balloon surface is covered
with RF reflective layer, which need be only several skin depths thick, of
order 10-20~$\mu$m at the wavelengths of interest.
Although only a fraction of the physical area of the reflective region 
contributes to radio signal collection from any given direction, this
area is still of order 100 square meters or more, equivalent to an 11 m diameter
single-dish for 360 degrees of azimuth, and a usable range of 10-15 degrees
in elevation angle.

Because we intend to operate
at meter-scale wavelengths (from 0.5-2 m, or 600-150 MHz in frequency)  
where the Antarctic ice is nearly transparent, and galactic radio
backgrounds are not an issue, and because we do not require diffraction-limited
optics for mission success, we find that the balloon surface is quite acceptable as
a focusing system. The location of the prime focus for the toroidal geometry
is on an inner toroidal surface within the outer balloon. A planar-patch
antenna array will be mounted on the surface of an inner balloon, with a
grid spacing that will be optimized for cost-effective detection of
the neutrino signals at a sensitivity that is at least an order of
magnitude, and in the best case two orders of magnitude, better than any
previous experiment. Triggering and data recording will be done on a conventional
payload platform at the typical payload location at the base of the balloon
system. Signals will be transferred to this package via analog optical fiber
transceivers.

As we have noted above, EVA's heritage is loosely based on the ANITA payload,
the only prior experiment to search for UHE neutrinos via a long-duration
balloon flight. ANITA currently has the best world neutrino limits in the EeV
energy range~\cite{ANITA2}. The strength of these limits derives from a key 
energy-dependent parameter of the detector: the effective acceptance of
the neutrino aperture as a function of energy. Since we are concerned
with an isotropic flux of neutrinos, it is the ``all-sky'' sensitivity
that matters, and in ANITA's case such sensitivity was achieved by
an array of 40 horn antennas with angular response functions that
each covered a patch of order 40 degrees in diameter.
ANITA's antennas were pointed at an angle of about 4 degrees below the
horizon (the horizon is about 6 degrees below the horizontal at typical
balloon float altitudes of $\sim 37$~km ) 
to maximize the volume of ice in their view, and their response thus
extended to about 24 degrees below the horizon, covering more than 99\% of
the area in view. A combination of 4 or more antennas with overlapping response functions
then determined a trigger when a radio impulse exceeded a pre-determined
level above thermal noise. 

For ANITA the neutrino acceptance was determined by
how far out toward the horizon the system could trigger on a neutrino-generated
radio impulse of a given radio intensity determined to first order 
by the neutrino energy. ANITA's threshold for triggering on a radio signal was
limited primarily only by the effective collecting area of the antennas involved.
In practice, this effective area was of order 1 square meter. EVA will increase
this effective area by a factor of 100 or more, and since 
the coherent Askaryan radiation
total power increases as the square of shower energy, a factor of 100 increase in
collecting area is required to get a factor of 10 improvement in neutrino energy
threshold.

\subsubsection*{Energy Threshold \& Sensitivity.}

The sensitivity of a radio antenna and receiver is determined by its 
collecting area, the receiver's bandwidth and integration time, and by 
the thermal noise background, which is the sum of the received thermal
noise power from blackbody emission in the antenna field-of-view 
(also known as its ``main beam'') and additive thermal noise from
components such as amplifiers, cables, and filters that comprise the
receiver system downstream of the antenna. The two contributions are
often expressed as a sum of the antenna temperature $T_{ant}$
and the system temperature $T_{sys}$.

For a system designed to record only the impulsive events some type
of threshold-crossing trigger is used to signal the presence of
a signal peak in excess of background fluctuations due to thermal
noise. The rms level of receiver 
voltage fluctuations in this thermal noise is given by
\begin{equation}
V_{rms} = \sqrt{k(T_{sys}+T_{ant})Z \Delta \nu }
\label{vrms}
\end{equation}
where $k$ is Boltzmann's constant, $Z$ is the receiver impedance,
and $\Delta \nu$ the bandwidth. The antenna temperature 
is an average of the radiation temperature of all objects in
the field of view, weighted by their apparent solid angle within the main
beam (or main region of angular response) of the antenna. 
In EVA's case $T_{ant}$ will be an average of the ice
temperature, of order 230~K, and the sky temperature, which
is several Kelvin at frequencies above 200~MHz. Noise figures
of current off-the-shelf low-noise amplifiers are typically
90~K or less.
Based on our ANITA experience with the Antarctic thermal noise environment,
the root-mean-square receiver voltage for a 500~MHz
band will be of order 10~microvolts, referenced to the LNA input, based on
equation~\ref{vrms} above, using $T_{sys} \sim 300$~K.

\begin{figure}[!htb]
\centering
\includegraphics[width=2.75in]{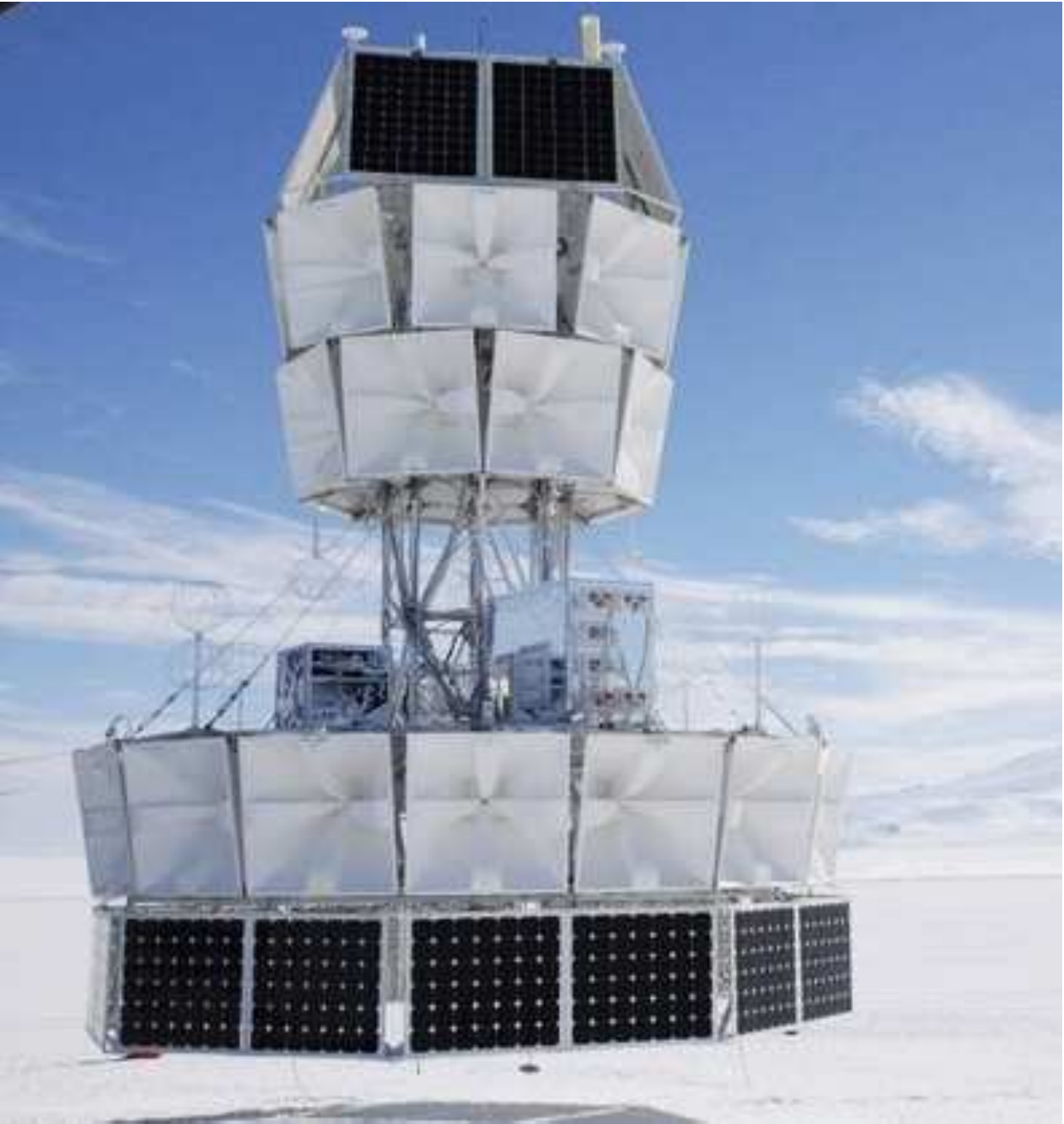}
\begin{small}
\caption{\it ANITA payload in flight-ready configuration.\label{payload06a}}
\end{small}
\end{figure}

\paragraph*{Antenna effective area}
ANITA's quad-ridged horn antennas (seen in Fig.~\ref{payload06a})
determine to a large part the overall sensitivity, or its ability to
``trigger'' on RF impulses.  These antennas had a
boresight directivity gain of $10\log_{10}(G)\simeq 10$~dBi 
(dBi = decibels relative to
an isotropic antenna), and it was roughly
constant over ANITA's frequency range of 0.2-1.2 GHz. 
Antenna effective collecting 
area $A_{eff}$ in terms of directivity gain $G$
and frequency $\nu$ 
is given by 
\begin{equation}
A_{eff} = Gc^2/(4\pi\nu^2)
\label{gain-eq}
\end{equation}
where $c$ is the speed of light. Thus
an antenna with constant gain vs. frequency has a
collecting area which decreases as the square of
increasing frequency. In practice, it was the off-boresight ANITA antenna
gain at the typical adjacent-horn overlap angle for
any given impulse that determines the average
effective area for neutrino impulses, and this 
reduces the gain down to about 7 dBi . Since ANITA required a
broad-band frequency content to trigger, the 
effective center-frequency at which to evaluate the
average gain is about 400~MHz, and the implied
effective area is 0.22~m$^2$ per antenna. In forming
a trigger, the signals from 4 antennas that can view
an incoming plane wave are used in a threshold-crossing
square-law discriminator, but are not combined
coherently, so the improvement in combinatorics
goes as the square root of the number of combined
antennas, and the net effective area for ANITA, in
determining the radio sensitivity, is of order 0.5~m$^2$.
This corresponds to an effective gain of
$G_{eff} = 4\pi A_{eff} / \lambda^2 \simeq 11.2 = 10.5$~dBi.

This effective area sets the scale of the current
state of the art for a balloon-based neutrino detector.
For EVA our design requirement is to thus increase
the effective area of the antenna by at least an
order of magnitude, with a design goal of two orders of 
magnitude, a factor of 100. 
The implied minimum and desired gains,
using again 400~MHz as the 
reference frequency, are
\begin{eqnarray}
G_{min} &=& 112 = 20.5~{\rm dBi} \nonumber \\
G_{goal} &=& 1120 = 30.5~{\rm dBi} \nonumber
\end{eqnarray}
above an isotropic antenna. An estimate
of the equivalent circular aperture dish diameter $D$ required 
to achieve this gain can be made by the
approximation $$ D = \sqrt{\frac{\lambda^2 G}{\pi}} $$
which indicates equivalent diameters of $D=4.5,~14$~m for these two cases.

\subsubsection*{Toroidal Balloon Reflector}

Radio reflector antennas based on a toroidal geometry
were first described by Kelleher and Hibbs in 1953~\cite{Kelleher53}, 
and Peeler and Archer in 1954~\cite{Peeler54}
and summary analyses may be found in several modern
antenna textbooks~\cite{AE,Wolff}. The designs considered in
prior work utilized a parabolic or elliptical curve rotated around an
axis in its plane to form a surface of revolution. Although 
only those surface generated by rotating
a circular arc actually conform to a true toroidal shape, all of
these reflectors have been termed toroidal reflectors; we will
follow this for convenience.

Here we choose the
axis of rotation to be the vertical $z$-axis, and thus the
reflective surface extends in the $\phi$, or azimuthal,  direction.
Obviously such surfaces of rotation must occupy only a portion of a full circle
of rotation or incoming radiation would be blocked either fully or
partially. Use of off-axis parabolic curves does not help in
this case since the incoming rays are paraxial and thus,
although the feed region does not occult the incoming wave,
other portions of the reflector surface will, if they extend
far enough around in azimuth. This constraint is acceptable for
many designs involving toroidal reflectors, since they may
still scan a wide range of azimuth, many tens of degrees or more.

However, in our case, the design goal is a system that can
scan $360^{\circ}$ in azimuth, and a limited range of elevation angle , of
order $10^{\circ}$. Use of a parabolic section
for the surface of rotation would in this case give strong
aberrations if used well off the main parabolic axis, so we 
consider a new configuration of the toroidal reflector with
a near-circular generating curve, rotated around the $z$-axis
to produce a complete toroid. Incoming plane waves enter
below the reflective band, and are focused to a region
above it. In our application, the incoming wave will be an
RF impulse, and it will enter the toroid below the
lower rim opposite to the active focusing area that will
apply for that direction. Once reaching the opposite side,
it is focused into an off-axis location with geometry
that is locally an off-axis segment of a spheroidal
mirror. 

At the focal plane, a set of feed antennas,
which are flexible planar patch antennas affixed to the
surface of an inner balloon of appropriate size to
approximately match the focal plane angle, receive the
focused plane wave. The polyethylene surface and load tendon
material of the balloon play no role in any of the wave
propagation, as these materials are completely transparent
and extremely thin on the scale of the wavelengths of 
interest here. The
generating curve is determined by the average free-surface of the
balloon, and such surfaces have been the subject of much
detailed theoretical and 
experimental investigation~\cite{Baginski1, Baginski2}.
This design will retain spherical aberration, as noted
previously, but such aberrations are generally more tolerable
than the extremes of coma that develop in 
short-focal-ratio off-axis parabolic systems.

In practice a large scientific balloon requires vertical load 
tendons to support the total weight of the system, and these
introduce a lobed structure for the balloon, leading to
a scalloped surface. We will account for the scallops in our
modeling of the surface, and they do constrain the highest
radio frequency at which a balloon surface may be considered
a coherent reflector.

\paragraph*{Zero-pressure vs. Super-pressure balloons.}
Stratospheric balloons have traditionally been constructed and
filled such that they reach their full inflation at 
altitude at an equilibrium pressure equal to the
ambient pressure at the float altitude, thus yielding
a zero-pressure offset from the surrounding atmosphere.
Vents on the base of the balloon help to ensure that
any over-pressuring is avoided. Such balloons are known as
Zero-Pressure Balloons (ZPB). The major drawback of ZPBs for
our application is the fact that their shape can change 
dramatically depending on their thermal environment~\cite{Baginski1};
for example, in Antarctica in the austral summer, cooling of
our ANITA-2 ZPB while over the very cold East Antarctic
Ice Sheet caused a decrease of order 40\% in the volume
of that 29 Mcft ZPB, and a loss of altitude of 123,000 to
113,000 ft. In an Appendix, we provide a detailed 
description of the analytical basis for the two different
balloon surface geometries.

After investigating to what degree we could tolerate such changes,
we concluded that the preferable vehicle for our
proposed application is a Super-Pressure Balloon (SPB)
such as those that are now being developed for the 
NASA SPB Program~\cite{Baginski1,Baginski2}. 
We  note that
the ``pumpkin'' shape of the current SPB design, with a ratio
of the polar ($R_p$) to equatorial radius ($R_e$) of 
$R_p/R_e \simeq 0.6$, does give
a larger astigmatism than a spherical design with more equal radii of
curvature, and 
the physical optics of such systems would improve with a larger
ratio $R_p/R_e$. But as we find in the next section, our
initial investigations of such balloons as reflector systems is
very promising.

\begin{figure}[htb!]
\centering
\includegraphics[width=3.15in]{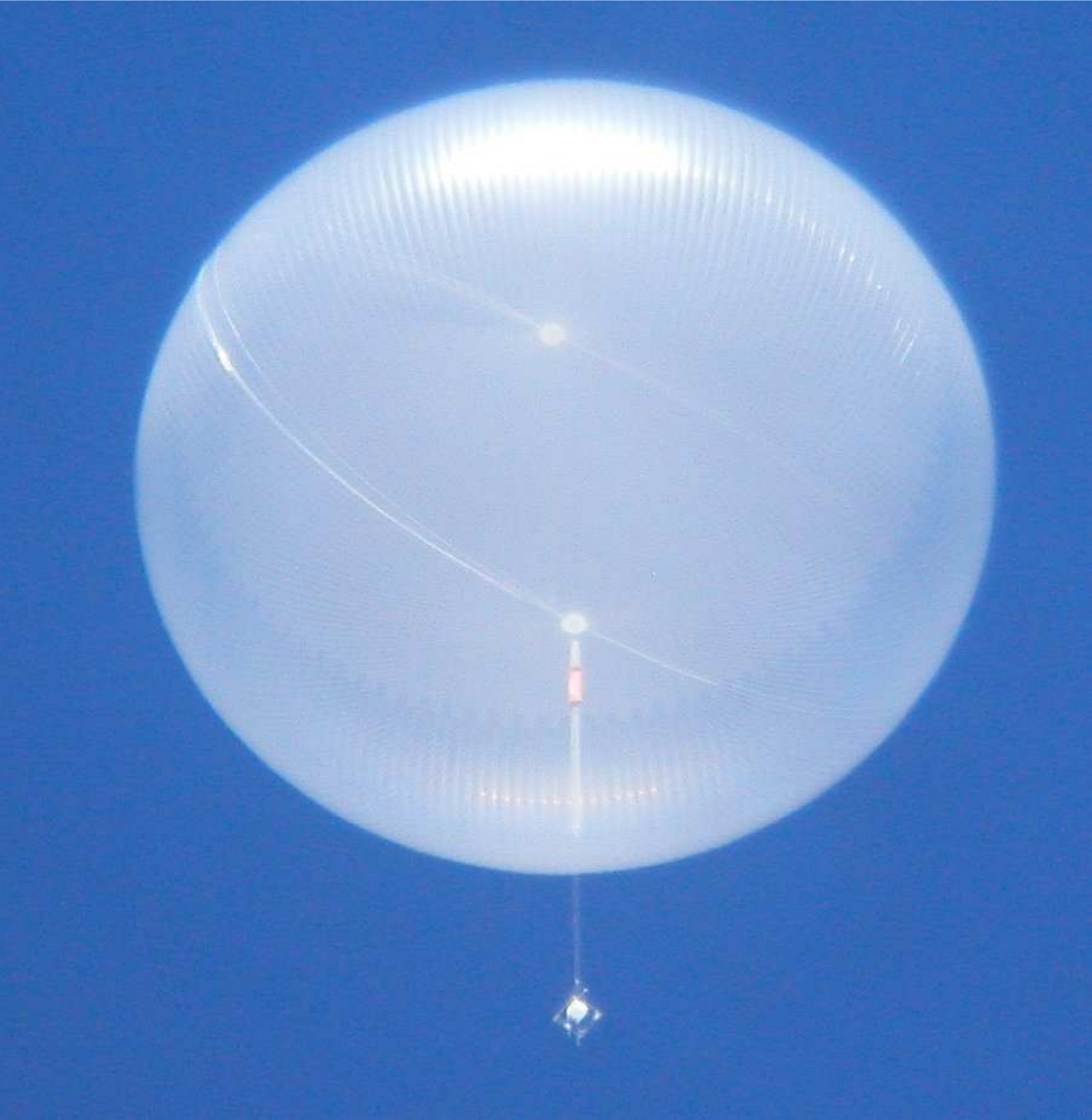}
\begin{small}
\caption{ Flight 591NT SPB at float altitude during the 2009 flight.
\label{SPB}}
\end{small}
\end{figure}

\subsubsection*{Analytical design guides.}

To select an initial geometry for the reflective section of the
balloon for numerical modeling purposes, we refer to results derived
for toroidal reflectors with parabolic generating curves as a guide.
Adapting these results for a spherical generating curve, 
the path difference function for any location on the
reflector surface can be approximated analytically with
first-order accuracy, following Wolff~(1966)~\cite{Wolff} as
\begin{equation}
\Delta = -2\hat{f}\left [1-  \frac{\cos \phi' ( 2 - \zeta -\cos{\phi'} + \hat{f}\cos \phi'}{1 + \hat{f} - \zeta } \right ]
\end{equation}
where $\hat{f}$ is the effective focal length normalized to a unit equatorial radius for the 
balloon (for spherical surfaces the physical focal length $f \sim R/2$); $\phi'$ is the
azimuthal angle of the location of the reflection point relative to the direction of the
incoming plane wave, $\zeta = \sqrt{1-z^2}$, 
and $z$ is the vertical coordinate, normalized to the equatorial radius.

\begin{figure}[htb!]
\centering
\includegraphics[width=3.15in]{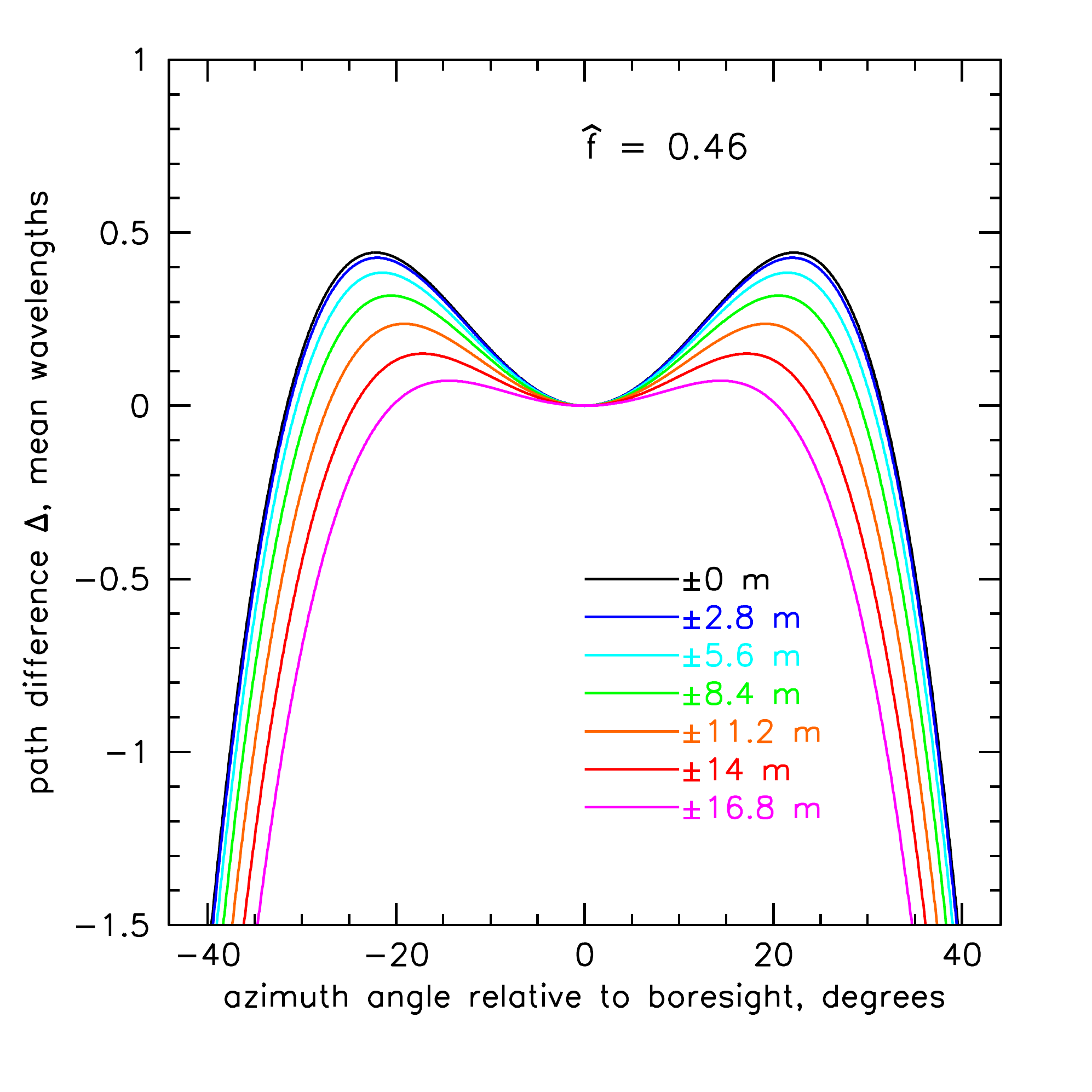}
\begin{small}
\caption{ Path difference of toroidal reflector for a normalized focal distance of
$\hat{f}=0.46$, for various slices of elevation above the horizontal mid-plane of the
balloon.
\label{ToroidDelta}}
\end{small}
\end{figure}

To simplify the numerical modeling requirements, we wish to simulate only an azimuthal 
subregion of the total
toroidal band, where the azimuth we are considering in this case is not the azimuth of an incoming
plane wave, but that of the toroid as viewed from the center of the balloon. 
The overall reflection geometry ensures that the azimuthal width of the appropriate 
subregion is well under $\pm 90^{\circ}$
around the optical axis, but to refine this we use the 
analytic results to determine an acceptable subregion. 
A family of curves, showing the path difference for reflected rays at slices of equal elevation 
derived from this equation is shown in Fig.~\ref{ToroidDelta}, where
in this case $\hat{f}=0.46$ and the radius
of the balloon here is 56~m. It is evident that at distances off the equatorial plane,
the phase error, while improving at higher $z$ at low azimuth angles, rapidly degrades
at higher azimuth angles. Using the Rayleigh quarter-wave criterion as a guide, it
is evident that the azimuth region over which the phase difference is acceptable extends to about
$\pm 30^{\circ}$ for a reflective band of order 20~m high, giving of order 1000~m$^2$ of
usable coherent surface area in the ideal case. In practice we will restrict the
simulation to smaller subregions of the reflector, because of other constraints
such as the clear aperture for incoming plane waves, but these results provide good
overall guidance for developing the models.

These results also apply only to on-axis plane-waves, and our design will also require 
focusing of rays that are well off-axis. Recent results in analyzing parabolic toroidal reflectors
have validated elevation scanning ranges of $>7^{\circ}$~\cite{Chu89} with
less than 1~dB reduction in gain, and azimuthal ranges for toroidal reflectors have
been used effectively out to the $\pm 30^{\circ}$ suggested here, or even more. For
purposes of our investigation, we consider initially
a reflective band of 11~m total height, and  about 50~m width, corresponding to an
azimuthal range of $\pm 25^{\circ}$ for a 56~m radius balloon. 

As we will see below, the standard balloon surface has 
a factor-of-two different radii of curvature in the circumferential and meridional
directions, and this will have important impact on the behavior of the 
surface as a focusing system. Such a surface is too complex to easily quantify
within the analytic framework, here but it provides us with a starting design which
we will evaluate numerically below.

A photograph of the fully inflated 7 Mcft superpressure balloon used in
the successful 54 day flight 591NT can be seen in
Fig.~\ref{SPB}. In early 2011, a successful  Antarctic flight of a 14Mcft design
was completed (flight 616NT~\cite{Cathey2011}), 
verifying the scaling criteria used to develop this
next step in the program. To date, the most difficult problem associated
with each step in scaling up the SPB designs has been the deployment of the
balloon as it inflates. In earlier designs, lobes would often remain
folded over one another even in a fully inflated balloon, leading to
distortion and unplanned stress in the balloon surface. However, in the
most recent designs, this problem appears to have been solved through
a new analytic method~\cite{Baginski3}, and all recent SPB launches have
deployed successfully.

\subsubsection{SPB shape stability}

Although SPB shapes are clearly far more stable than ZPB shapes, there are
still variations in the differential pressure of the inflated balloon that
could lead to changes in the focal properties of the surface. Typical mean
differential overpressure for the SPB tests to date are about 50-60 Pa,
and variations of $\pm 30-40$\% have been observed due to variations in the
solar angle and float environment.
Baginski \& Brakke~\cite{Baginski3} 
considered the strained shape of the gore for the Flight 591-NT design, which also
formed the basis for the 616NT designs as well. 
Let H be the height of the balloon (measured from nadir fitting to apex fitting) and D to be
the diameter. They found  $H = 50.32$~m and $D = 82.11$~m when the nadir differential pressure 
$P_0 = 120$~Pa.  In Table~\ref{focal}, below we present the dimensions 
when  $P_0$ is between 19.9 and 120 Pa;
$H$ and $D$ were not reported in \cite{Baginski3}.

\begin{table}[hbt!]
\renewcommand{\baselinestretch}{1}
\begin{center}
\caption{\small \it Super-pressure balloon height and diameter as a function of
differential balloon pressure, as found in~\cite{Baginski3}, for flight 591NT.
\label{focal}}
\vspace{3mm}
  \begin{tabular}{|ccc|}
\hline 
$P_0$  (Pa)  & $H$(m)     &   $D$(m) \\ \hline
 120.500  &   50.325  & 82.115  \\  
 99.256   &   50.326  & 82.021  \\
 77.991  &    50.355  & 81.920  \\
 56.867  &    50.444  & 81.804  \\
 36.522   &   50.674  & 81.654  \\
 19.902   &   51.248  & 81.416 \\ \hline
\end{tabular}
\end{center}
\end{table}

For the actual flight, pressure gauges indicated that $P_0$  varied between a minimum of
30 Pa (night) and a maximum of 75 Pa (day) (see~\cite{Cathey}). 
From the third and sixth entries in Table~\ref{focal},
we find $\triangle D= 0.5$~m and $|\triangle H| = 0.9$~m. Similar variations were
seen for the 14 Mcft balloon in flight 616NT~\cite{Cathey2011}.
Our analysis indicates that such changes will lead to focal effects that are well
within the depth-of-focus tolerance ranges we have found both in simulation and
our scale-model tests, as we report below.

In addition to pressure-induced shape changes, any pendulum motions of the
balloon could also have an impact on the stability of the field-of-view of the
reflector surface. Fortunately, pendulum motion of stratospheric balloons at float
has been measured for many different flights. Apart from a period of a few hours once
the balloon reaches its float altitude, pendulum-induced tilts in the balloon are
very small, a small fraction of a degree typically.

\subsubsection*{NEC2 Modeling.}
As a proof-of-concept we have created a detailed
antenna model for a geometry involving a 29 Mcft SPB balloon
with approximately 56 m radius, and this model is the basis
for several of the figures provided here. This balloon size
presents a significant advance in the current scale of these balloons,
but is a standard ZPB size, and there appear to be no technological 
limitations to scaling up the current SPB designs to this level.
However, we note that after development of this model and the
analysis described here, the SPB program adopted a goal
of a 25~Mcft design for future use, and our scale models 
currently built and planned will all conform to this design.

Antenna modeling is done with the 
Numerical Electromagnetics Code version 2 (NEC2)~\cite{NEC2},
a method-of-moments full-electromagnetic solver based
primarily on wire-frame modeling of conductive 
structures. It has the virtue of allowing straightforward
wire-frame model prototypes to be built up into fairly complex
structures, which can produce realistic results even for
surface modeling, although they will underestimate
reflectivity of continuous conductor 
surfaces because of the wire grid approximation. 
Although NEC2 can also use facet patch
segments to create filled surfaces, this capability of the
code is much harder to accurately utilize, and thus we have made
use of only wire-grid models here.

\begin{figure}[htb!]
\centering
\includegraphics[width=3.1in]{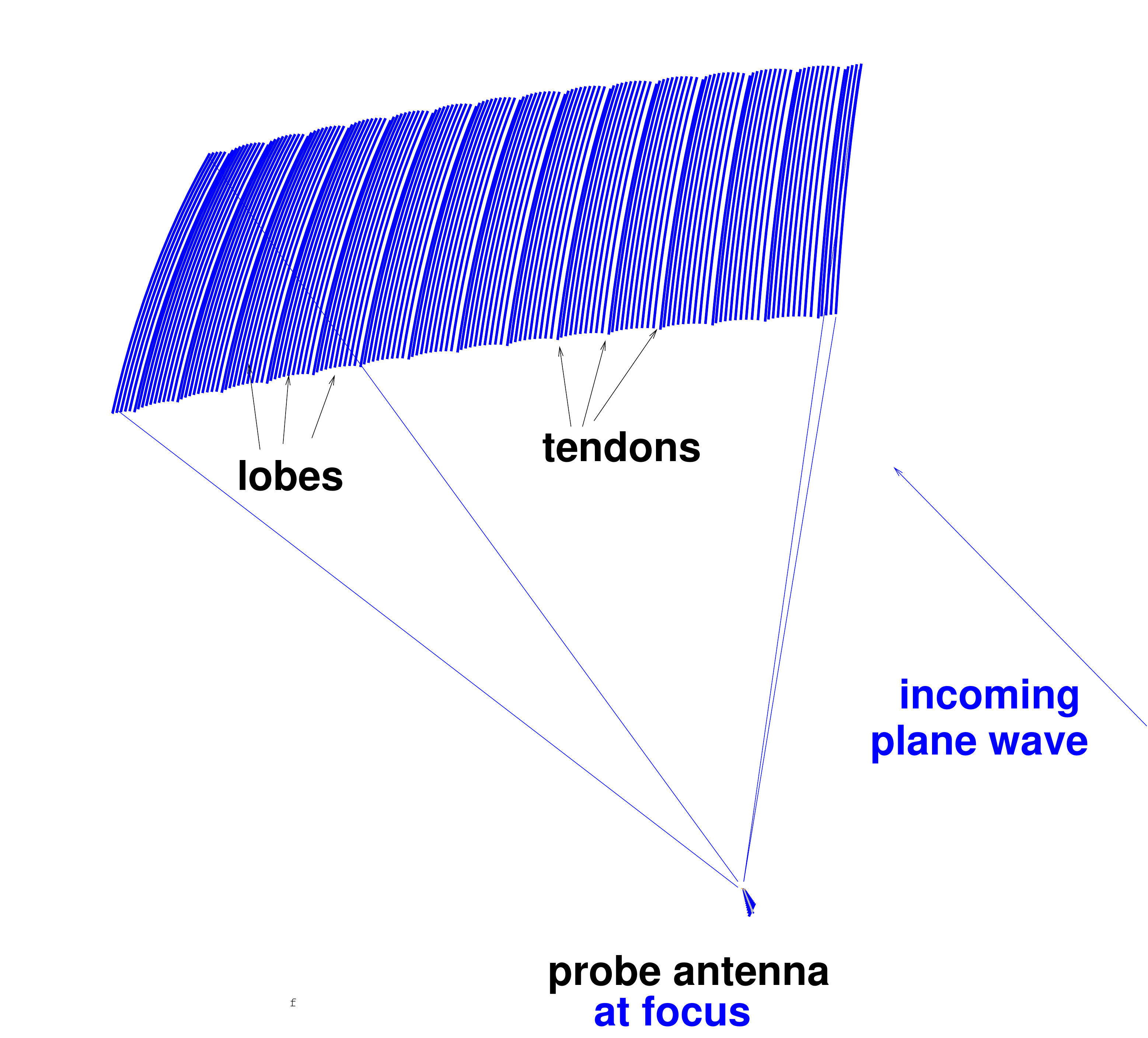}
\caption{ \it \footnotesize{NEC2 model geometry used for gain estimation.
\label{nec2model}}}
\end{figure}

A detail of the model is shown in Fig.~\ref{nec2model}. 
In practice the NEC2 models which use the entire array are prohibitively
large for calculations, and as we have noted above,
we thus restrict the modeling to a 
subregion of the reflector approximately $\pm 25$ degrees wide in
azimuth and 11~m high; these choices were deemed conservative given the
additional constraints we have that go beyond the analytic model.
This section is shown in the Figure, along
with the location of the wideband probe antenna which is used to excite
the structure in the NEC2 model. All of the modeling was done using
the reciprocity relations for antennas, which ensure that 
transmitted-wave response functions may be used to estimate the received-wave
response functions. 

We model the surface as having two 
radii of curvature: the circumferential, or ``hoop'' radius $R_h$, and
the meridional radius of curvature $R_m$. the relation $R_h = 2R_m$
is generally true for pumpkin balloons due to the intrinsic shape
of the {\it Euler-elastica}, which is the analytically derived curve 
for such surfaces. 

There are thus two primary aberrations involved in
the image formation of these reflector segments: (1) Spherical
Aberration, which causes central rays near the optic axis to
focus further away than the marginal rays on the outer edges of
the reflector; and (2) Astigmatism, which causes a different focal
length for the two different radii of curvature. 

In addition to these primary aberrations, we have also accounted for
a secondary aberration which is important to balloon surfaces:
the tendon-gore structure of balloon fabrication leads to a
scalloped surface where the tendons have a smaller radius than the
center of the gore, and lobes form in the surface. 
In addition to the primary radii of curvature $R_m$
and $R_h$, the individual pumpkin lobes have a bulge radius  of $r_b$,
typically several meters, as described in the Appendix.

\begin{figure*}[htb!]
\centering
\includegraphics[width=6.5in]{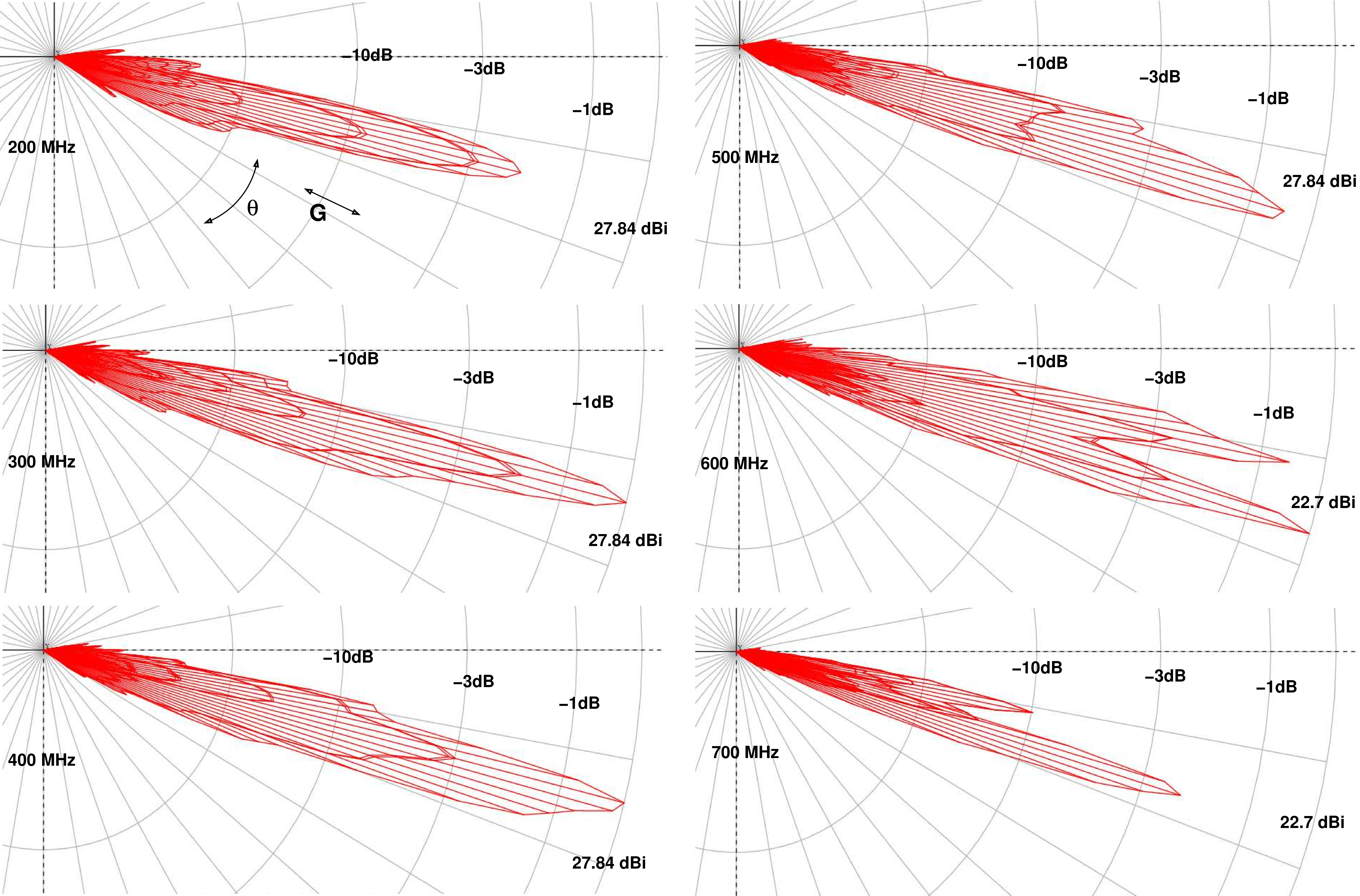}
\begin{small}
\caption{ Elevation gain polar plots for six
frequencies and the largest off-axis elevation angle
(about -13 degrees) expected to be observed
with EVA. The gain peaks at about 27.8 dBi,
exceeding the design goal of 26.9 dBi indicated
previously. Note the gain scale change in the two
highest frequencies.
\label{SPBelgain}}
\end{small}
\end{figure*}

For our 
simulated balloon we
have assumed about 300 total gores, several of
which are evident in Fig.~\ref{nec2model}.
We have assumed a typical case where the fully-inflated lobe center is 
about 10 cm further out than a line joining the
tendons. This value is about 20\% of the wavelength at 600~MHz which is 
currently the upper end of our design frequency.

\subsubsection{Model results.}

The model NEC2 analysis was done by iterative optimization of
the gain vs. frequency and angle for different locations of
a probe antenna which acted as the feed excitation for the
larger toroidal section. Starting values for the best feed
location were based on the sagittal and transverse foci of
the astigmatic reflector. We found that the best gain values and
apparent circle of least confusion for the focused antenna beam
tended to favor that equatorial radius of curvature, primarily
because there was more reflective area available at the larger
focal distance, and the depth of focus was also larger. We did
not adjust the width of the vertical reflective section; this
is another parameter that could be optimized. Our NEC2 model
filled the aperture of the reflector with the equivalent of
10~cm diameter ``wires'' at a spacing of about 35 cm, 
corresponding naively to about 30\% reflectivity of the surface;
this value was chosen as the closest that we could 
place the elements while still maintaining robust numerical
stability for the calculation. Thus the estimated
antenna gains should show some improvement with
fully-filled reflective sections.

\begin{figure}[htb!]
\centering
\includegraphics[width=3.3in]{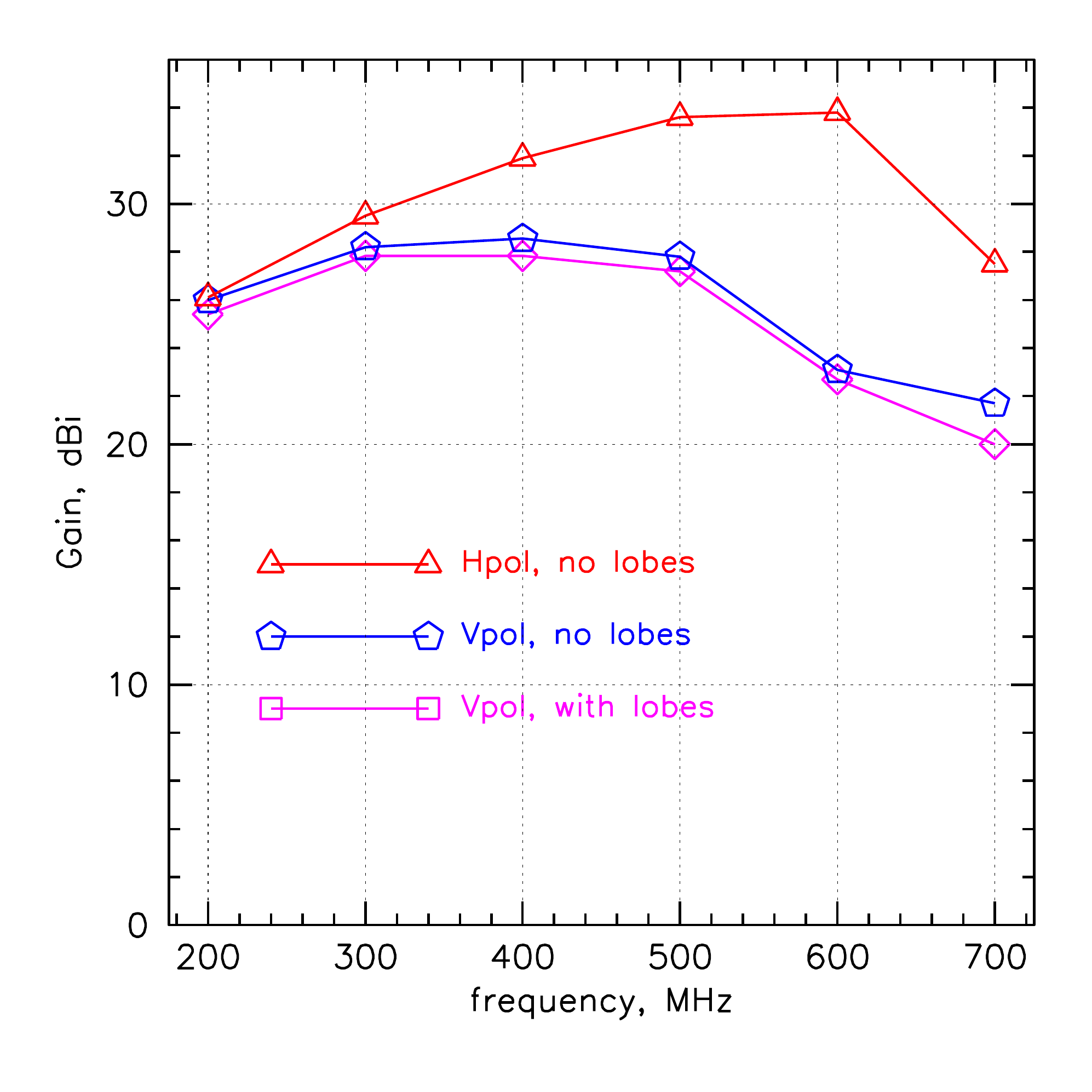}
\begin{small}
\caption{ Peak gain vs. frequency for the models described in the
text, including vertical polarization with and without the lobe-tendon
structure of the surface, and horizontal polarization, also with a smooth
surface here.
\label{EVA_gain1}}
\end{small}
\end{figure}

Fig.~\ref{SPBelgain} shows the primary results of the simulation: polar
plots of the antenna absolute directivity gain vs. elevation angle for
the largest off-axis elevation angles of interest for EVA. 
The toroidal antenna is designed to observe the angular range 
from a few degrees above the geoidal horizon (which is about 6.8 degrees below
horizontal for our suborbital altitude) to an angle of about 8 degrees below the horizon,
a range which covers 96\% of the Earth's surface area in view in the lower hemisphere. 
Event plane-waves that arrive from
steeper upcoming angles are reflected progressively further off the
optic axis, and are thus subject to more sever aberrations.
The results shown here are for the one of the steeper angles of
interest, $13^{\circ}$ below the horizontal.
We have also confirmed that the shallower angles corresponding
to events near the horizon do give better results and slightly higher
gain.

The plots show a wireframe structure of the gain values, where
elevation is sampled in 1 degree increments. 
Azimuth angle structure can be derived also from this plot by noting
the location of the transverse wire frame contours of the 
azimuth sampling, which was done here
in 0.5 degree increments .
The plots cover the range from 200-700~MHz, and the gain over this
range has a broad plateau of 26-28 dBi from 200-500~MHz, and then
drops to 22.7 and 20.4 dBi respectively at 600 and 700~MHz.

Clearly the image formation is seriously degraded at 600~MHz and above, 
with a double-lobed structure appearing in simulation at 600~MHz. It is
evident that the optical aberration at these higher frequencies has
grown to the point where partial destructive interference is taking place
in the image-forming region.
In any case, our 
results do indicate that over the frequency range from at least
200-500~MHz (a fractional bandwidth of order 1) the net directivity gain of the
system approaches within a factor of two of the design goal, and
already well above the minimum requirements. The antenna thus 
achieves a broadband response that is suprisingly good given that the
surface shape is not controlled in any way.

We have not yet attempted to further
quantify the cause of the high-frequency aberrations, nor have we
done any optimization of the system to improve the higher
frequency response.  We do note that it is relatively straightforward to include a
circumferential reflective membrane just within the outer inflation surface, which
can reduce the large difference between meridional and hoop 
curvature, as well as minimizing the effects of the lobes. While
we have not analyzed such a band in the current work, the potential for
improvement in bandwidth and gain could offset the additional complexity of
construction and deployment.

\paragraph{Variations: horizontal polarization, and suppressed lobe structure.}
The simulation results shown here are for the vertical
polarization with the lobe and tendon structure included in the model, as noted
above. We have also investigated the behavior of the model with the
lobe and tendon scalloping suppressed, as would be the case for a balloon
with reflective film that spanned the intergore regions. Such interior
membrane structures have also been mechanically modeled for their dynamics
on the balloon surface, and they appear to be benign in their behavior, with
respect to the balloon inflation and operation. 
In addition to this secondary model for vertical polarization, we also
developed a similar wireframe model for horizontal polarization to ensure that
there were no adverse effects in detecting a wide range of polarization planes.

We found that the lobe and tendon structure of the balloon surface has
very modest effects on the antenna gain, as shown in Fig.~\ref{EVA_gain1},
which plots the resulting gains vs. frequency for the three configurations
described here. We also found that the double-peak structure at 600~MHz
was present for vertical polarization even in the smooth-surface case, indicating
that the effect is likely due to the large scale aberrations rather than
the surface relief induced by the lobe structure. Given that the dual gain
peaks seen at 600 MHz are not centered on the gain peak at lower frequencies,
there is likely to be some destructive interference at that frequency which
is suppressing the main peak in favor of nearby sidelobes; there are techniques that
can remediate these effects, such as tapering of the reflectivity at the
top and bottom of the reflective band.

\begin{figure*}[htb!]
\centering
\centerline{\includegraphics[width=6.5in]{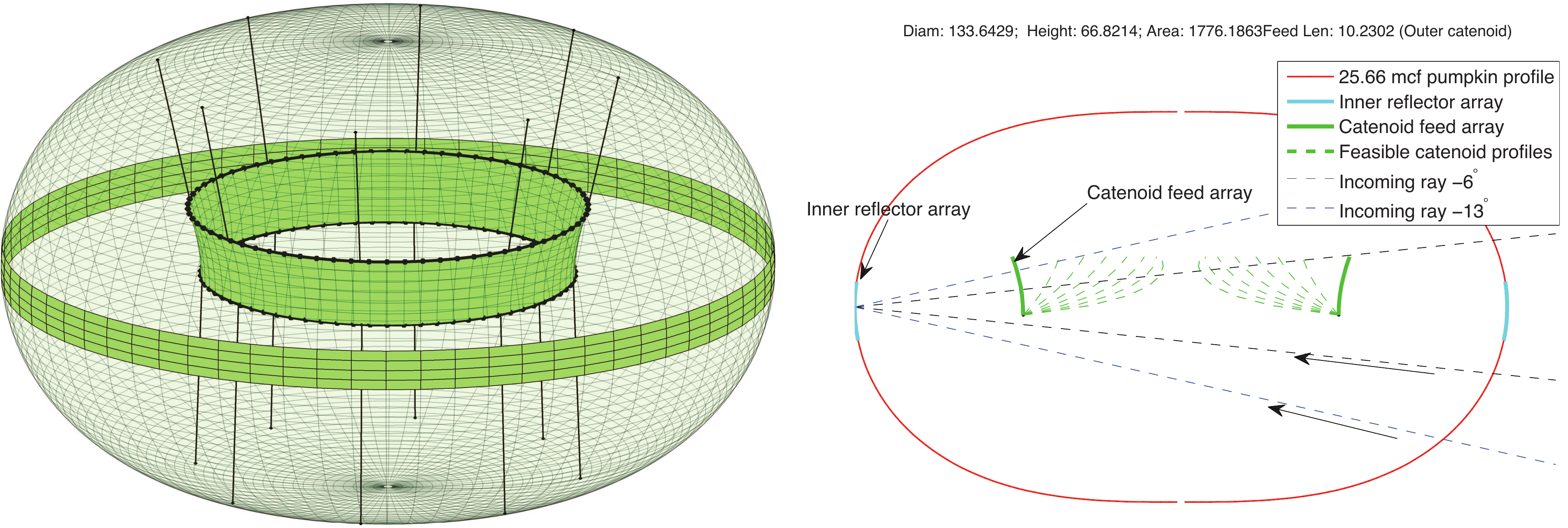}}
\caption{\it \footnotesize{Left: 3D model of a 25.66 Mcft superpressure balloon
with a possible catenoidal surface shown schematically as a location for the
feed array. Right: cross section of the balloon surface and catenoid, showing
the range of sections that are possible, and ray paths for the modeled
sections. }
\label{catenoid}}
\end{figure*}

Of equal interest is the fact that the horizontal polarization results show no
similar effects at 600~MHz, in fact we found that the gain curve retains a single
peak throughout this frequency range, and that the overall gain is several dB
higher, probably because of the lack of edge effects for the E-plane of the field
which is now aligned with the longer portion of the reflector. In any case,
the horizontal polarization performance exceeds that of the vertical polarization,
and well exceeds the requirements for total gain.

\subsubsection{Smaller SPB designs.}
All of the modeling done here for a balloon of
56~m radius can be applied to a smaller radius superpressure 
pumpkin with appropriate scaling. For example we also
investigated the 14.8 Mcft, 46~m radius SPB design, which was recently flown in
the 22-day NASA SPB flight 616NT~\cite{Cathey2011}. 
Under these conditions we
find that similar beam-pattern results are obtained 
if we scale the equatorial reflector
region height from 10~m down to 8.2~m. 
The expected gain in the mid-frequency region
at 300-400 MHz scales as $(D/\lambda)^2$ and we thus anticipate
a decrease of about 2 dB in overall gain, still preserving the
possibility that a smaller balloon may be able to achieve the
science goals for EVA, although the smaller free-lift capacity 
would levy restrictions on the
payload mass budget.

\subsubsection{Feed array system.}
The feed array system is currently envisioned to occupy an interior band around the
surface of an inner membrane. Lacking detailed knowledge of the shape of
the focal plane surface, we anticipate that the inner membrane
could either be over-pressured for a convex focal surface, or it may
be developed as a catenoid surface if a concave focal plane is necessary. Currently
our antenna modeling does not give sufficient information to clearly
delineate the focal surface structure; this will be determined as part of
the study we propose here. If the membrane is over-pressured in an upper
chamber to produce a convex surface, the design may be accommodated
with a membrane having cylindrical symmetry, fused to the upper surface,
and possibly tensioned to the central collar at the base of the balloon.
If, as appears more likely in our current models, the region of best-focus
is in fact concave, we have explored catenoid surfaces with a partial
membrane containing the feed array, as shown in Fig.~\ref{catenoid}.

Each feed antenna will be a low-gain planar patch antenna,
probably with dipole-like response. Normal radio astronomical systems
require that a feed antenna have a directivity that fills only its main
reflector, thus a fairly high gain (6-10 dBi) feed antenna may be required. 
In our case, since the antenna temperature of the main reflector will be
dominated by the 240~K temperature of the ice in its view, the feed antennas
need not be directed only at the main reflector aperture, as any ``spillover'' in
this case will be into either cold sky or ice. Also it is important to note 
that our feed antennas are not used as transmitters; a transmitter feed must not
waste power into directions other than its main reflector, but for a receiving
system, this consideration is irrelevant. 

For the focal length of our NEC2 model, which was of order $f=25$~m, the 
focal plane transverse scale is given to first order as $f^{-1}=0.04$ radians/m,
or about $2.5^{\circ}$ per meter. We need to sample over an elevation range of about
8 degrees, or just over 3~m wide. The broadband 
feed antennas are expected to be of order
$\lambda/2$ at the longest wavelength of interest, thus we expect about 
five antennas along the elevation scan direction. From the gain plots above
it is evident that the full-width-half-maximum of our elevation beam is
about $3-5^{\circ}$ in elevation, and about $0.5-1.5^{\circ}$ in azimuth. Thus
we anticipate oversampling in elevation, which will allow for multi-antenna
triggering as was used in ANITA, as well as phase-gradient measurements to
refine the elevation angle measurements of any event. 
Azimuth sampling will be matched to first order, giving angular resolution 
of $1-2^{\circ}$ in azimuth, which has proven to be adequate for ANITA. We
anticipate a total of 1200 patch antennas to cover the required solid angle.

\section{Results of microwave scale-model tests.}
\begin{figure*}[htb!]
\centering
\centerline{\includegraphics[width=6.5in]{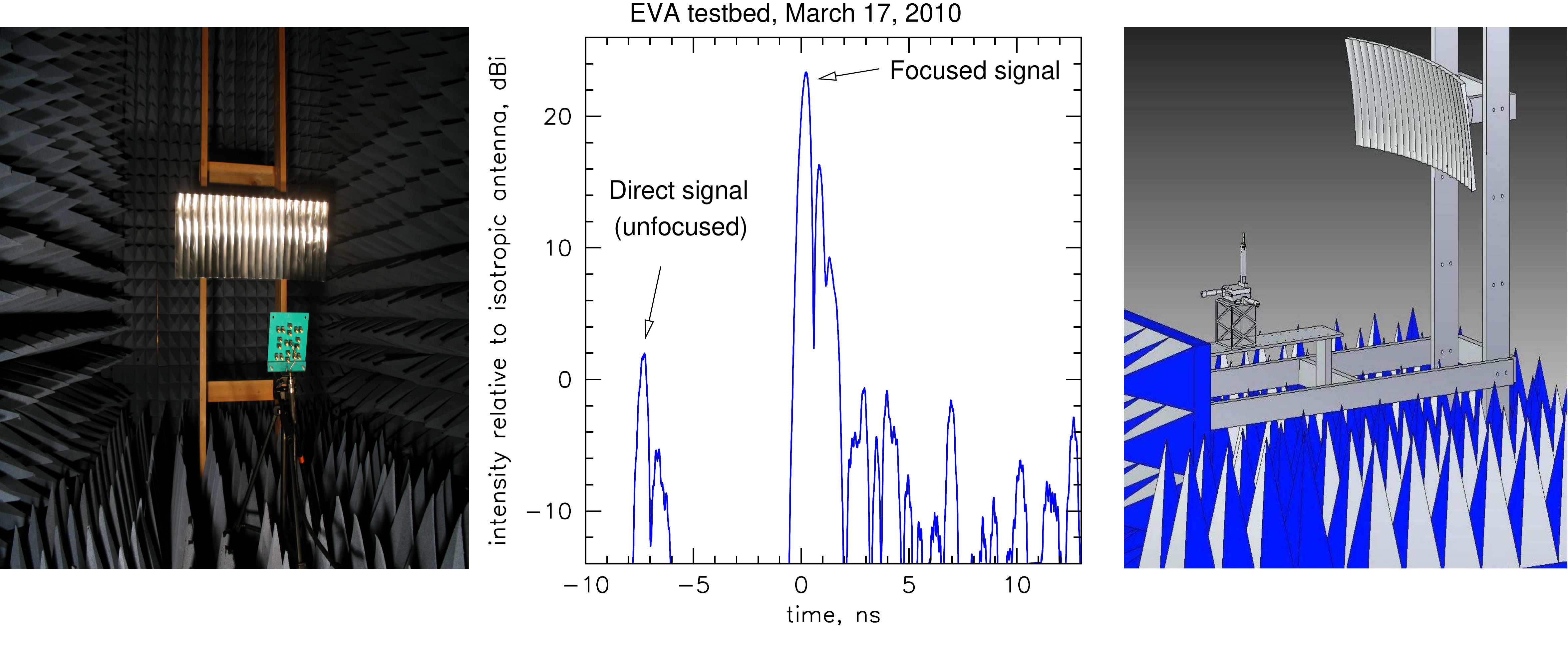}}
\caption{\it \footnotesize{Left: Photograph of 1/35 scale model of toroidal balloon section, equivalent
to a 10~m high, by 30~m wide section of the full balloon. 
Center: initial results of the focusing test
with the toroidal scale model, using a microwave pulse in the 6~GHz range (200-300~MHz equivalent scaled
frequency for the 1/26-scale model). 
The pulse to the left measures the unfocused incoming plane wave intensity using a 
patch dipole antenna, the much larger peak at center is the focused response with a peak
directivity gain of 23.4 dBi. Right: diagram of the test setup with the
1/26-scale model, including a micrometer stage at the focal region used to profile the focused
beam parameters.}
\label{EVAscale}}
\end{figure*}

We have constructed a microwave scale model testbed for exploring 
the behavior of sections of a toroidal surface, including the effects
of the balloon gores and tendons. This has been accomplished by computer-
numerically-controlled machining of exact 1/35 and 1/26 scale models of a 
computer generated model for the fully-deployed surface of a 25 Mcft
super-pressure balloon, following the current planning design adopted by
NASA for their SPB program. We have also developed several different
microwave receiver antennas, including a prototype patch array. A photograph
of one of the test setups using the 1/35-scale model, taken in an anechoic chamber 
is shown in Fig.~\ref{EVAscale}(Left). The scale model 
was illuminated with a 1.8~m wide collimated  microwave beam made by a large off-axis
paraboloidal reflector, situated opposite the EVA test stand. Because of constrants
of the scale model machining, we restricted the width of the reflective section to
a smaller region than was simulated in our NEC2 models; the scale model 
corresponds to a $\pm 15^{\circ}$-wide azimuthal section of the balloon rather
than the $\pm 25^{\circ}$ section that we simulated. As a result we
anticipated that we might observe somewhat lower total gain for the model vs. the
simulation.

Our initial impulse testing using the 1/35th scale model was challenging due to frequency
limitations imposed by our test equipment, and we could not accurately test the system
above an equivalent reference frequency of about 150~MHz. 
In our initial testing of this system, which was also 
limited by our current coarse positioning system, we measured a directivity gain of
order 16 dBi. Using what we learned from this model, we then constructed the larger 1/25-scale
model, again with near-perfect fidelity to the simulated balloon surface, as
afforded by the CNC system. We retained the lobes and tendons in our model to 
maintain a conservative approach, although it appears sxtraightforward to
suppress the effects of this structure 
in practice using internal reflective films that span the lobes.

With this new model, we have now measured a 
directivity gain factor in vertical polarization of up to
$G=220$ relative to an isotropic antenna, or in equivalent terms, 23.4 dBi, at a
microwave frequency of about 6.6~GHz, corresponding to 260~MHz for the full-scale. 
This value is within about 3 dB of the NEC2 model prediction, and already well above
our minimum performance requirement. Fig.~\ref{EVAscale} (left) shows a
photograph of one of the initial test setups, with a prototype patch array at the
focal plane. The middle pane of the same figure shows the layout of a micrometer stage
now in initial testing for higher-precision focal plane measurements. To the right
are the data validating the performance noted above. The smaller peak to the left is
the incoming plane wave impulse signal detected by the backlobe of the patch 
as it passes by before
focusing, then the much larger signal (here in logarithmic units of intensity) arrives
from the balloon reflector section. Due to constraints imposed by our transmitting system,
we have only been able to investigate vertically polarized signals to date, though
based on our simulation results, we expect that the horizontal polarization results
will likely exceed the vertically polarized results in their performance.

It is evident that these results already validate the
basic premise of EVA: that a toroidal reflective section on an unmodified super-pressure
balloon surface already possesses quite compelling optics for radio astrophysics
applications such as we envision.

\begin{figure}[htb!]
\centering
\centerline{\includegraphics[width=3.15in]{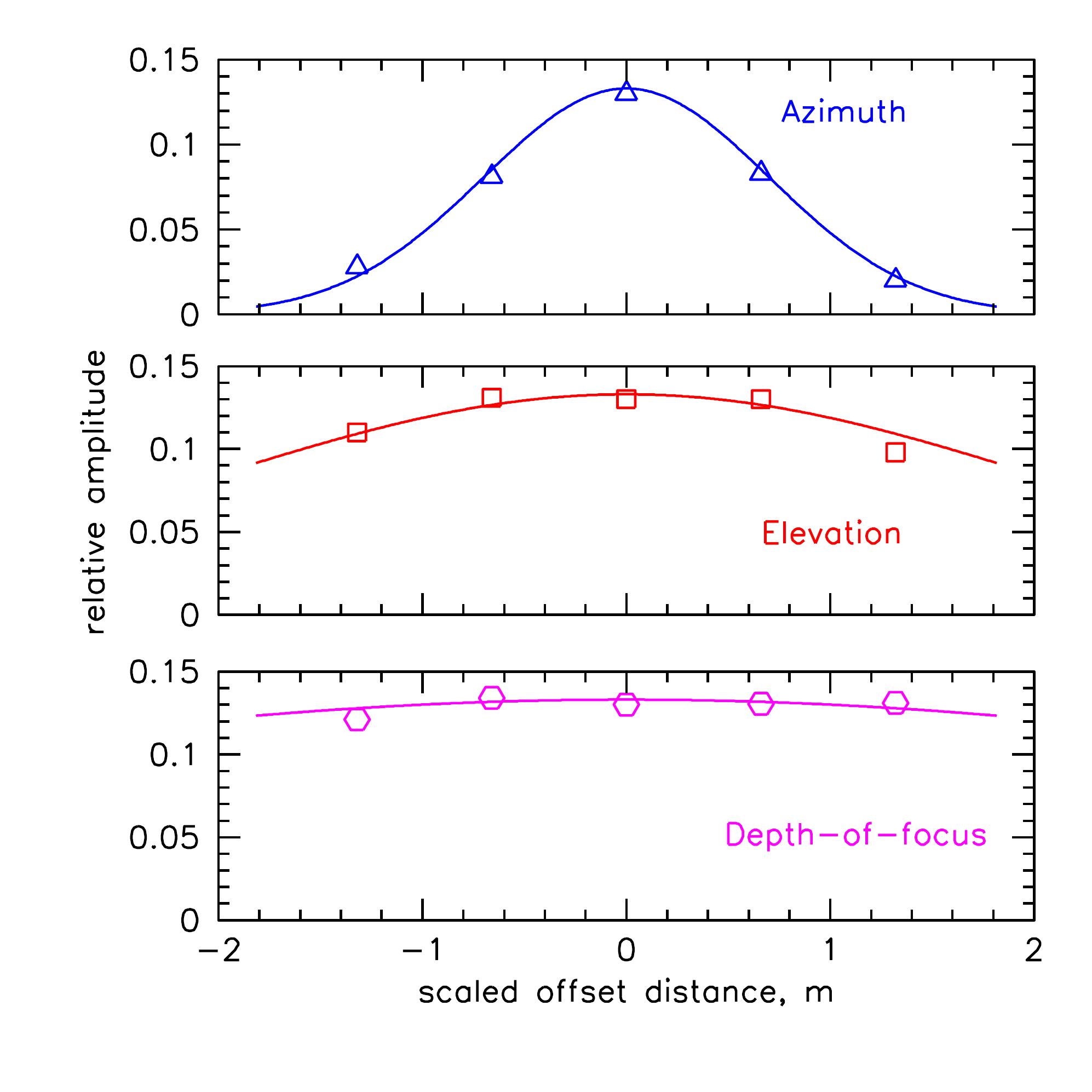}}
\caption{\it \footnotesize{ Three-dimensional focal region measurements (points) of the 
response function of the EVA model, with fitted
Gaussian beam curves shown. The dimensions have been scaled to the full size of
the reference balloon model.}
\label{EVAfocus}}
\end{figure}

Another important issue in EVA performance is the shape of the focused beam, including
the depth of focus, and the width and height of the main lobe of the beam in the focal plane.
This has important implications for the control fidelity of the membrane surface and
for the metrology precision. We have made preliminary measurements of the beam shape
over the principal axes on the 1/25th scale model. These measurements were made using
the micrometer system depicted in Fig.~\ref{EVAscale} (right), and the results are plotted
in Fig.~\ref{EVAfocus}, with the measurements (points) fitted to Gaussian beam parameters
(curves).

We find that
the beam is narrowest along its transverse dimension (eg, projected azimuth), with
a Full-Width at Half-Maximum (FWHM) of 1.14~m scaled equivalent. Along the direction of 
the beam, our micrometer translation was not adequate to find the half-power points, and
the resulting lower limit on the depth-of-focus is $> 3$~m. Transverse to the beam in the
vertical (eg. elevation) direction, we find again a very broad maximum with FWHM $\sim 3-4$~m.
These values are consistent with our modeling expectations, and provide us with clear
design direction for both the sampling density of the focal plane patch antenna array,
and for the level of control that will be necessary for the inner membrane in order
to retain good focal plane response.

\section{Other EVA Subsystems}

\subsection{Receivers.}
	
The RF front end for EVA will consist of an integral front-end bandpass filter followed by
a low-noise-amplifier (LNA)/power limiter combination, with about 36 dB of gain,
and a second stage amplifier designed to boost the signal up to the point where
an analog optical driver can modulate it onto optical fiber, an additional 20 dB or so.
These elements are all in close proximity to the patch antennas to ensure no
transfer losses through cables, and are enclosed in a flexible Faraday pouch for additional
EMI immunity. The signals from the LNA then drive an analog fiber-optic radio-frequency
link, and are transmitted via single-mode fiber along the inner balloon tendons, and
led down through the base collar to the flight train and eventually to the payload at the bottom.
The power budget for the single-patch LNA+2nd-stage amp+fiber 
transceiver is 0.8W per channel for 1200
patch antennas, and these are fed from the 1.2 kW photovoltaic array at 
the outer balloon-top platform (BTP). 
The expected noise performance of the LNA and bandpass filter combination is about 90~K on average
per channel, and we have already demonstrated this performance in the ANITA-2 flight.

\subsubsection{RF Interference}
The three completed flights of ANITA (and its prototype ANITA-lite) have also provided
excellent information regarding the anthropogenic noise environment in Antarctica,
and strategies for managing the receiver and trigger systems in the presence of such
noise~\cite{ANITA-inst}. ANITA has demonstrated that, apart from regions close to the
major bases -- McMurdo Station, South Pole Station, and a few others -- the
anthropogenic noise is sporadic and does not add significantly to the thermal noise
environment, nor does it saturate the trigger systems of instruments such as
ANITA and EVA. Clearly the higher sensitivity of EVA will lead to a higher trigger
rate on weak RFI, but such triggers can be rejected in real time by their association
with known bases or camps.

\subsection{Analog fiber-optic transceivers.}
Transmission of the data from the antenna feeds to the instrument
payload cannot be done via coaxial cable due to weight concerns,
therefore, RF over fiber technology will be used to convert RF to
optically modulated signal over single-mode fiber optic cable and
back to RF at the payload using a custom-built fiber interface
using commercially available laser diodes and PIN photodiode receivers.

The concept of using optical fiber to connect antennas at a distance
to the signal processing equipment ("antenna remoting") has been
actively studied by industry for WLAN distribution~\cite{Yakoob,Ackerman}
as well as several other scientific collaborations. Most recently, the
Australian Square Kilometer Array Pathfinder project conducted
experiments using a commercially available (\$30 in quantity) vertical
cavity surface-emitting laser (VCSEL) and demonstrated the feasibility
of connecting 7200 antenna feeds using directly-modulated VCSELs~\cite{Beresford}.
Improvements in high-efficiency VCSELs~\cite{Furukawa} should reduce the noise
figure to a tractable level. The primary power cost of the VCSEL
transmitter would be in the RF amplification needed to compensate
for the noise figure of the optical link, with a target of less than
0.5 W/channel -- about a factor of four below typical powers in current
commercial devices.

\subsection{Trigger/Digitizer System.}

We anticipate utilizing a trigger system closely based on ANITA heritage.
The incoming RF signal, once it has been extracted from the optical
receivers, is split conceptually into two paths, one of which is used for
triggering and the other for digitization of the waveform if a
trigger is detected. Because we are using a factor of 15 more signals
that ANITA, we will effectively implement a decimation scheme for 
digitizing the data, so that a trigger does not require digitization
of large portions of the feed array.

Improvements in the ASIC technology for low-power waveform samplers
have enabled deeper storage and demonstrated 
concurrent read/write operation~\cite{BLAB1}. Incorporation of a 
discriminator has also been shown to be feasible, and this allows
for the concurrent function of both trigger and sampling in a 
single ASIC. Thus while we still maintain the conceptual difference
of the triggering vs. digitization paths in our system, in practice,
the functions will be effectively co-located for EVA.


We have adopted the following preliminary specifications as a working
architecture for EVA.  Optical signals arriving
from the main E/O fiber bundle are broken into groups of 32 signals
each that go into approximately forty compact Sampling Trigger Modules
(STMs) which each contain Application Specific Integrated Circuits
(ASICs) based on a next-generation version of our current LABRADOR
chip~\cite{LAB}. These ASICs are designed to handle internally both
triggering and sampling via a switched-capacitor array. Digitization
only proceeds upon external command.

Local STM trigger signals are broadcast to a central DAQ module in the
system.  This trigger master than analyzes these trigger primitives
and broadcasts feed-array region-specific readout window requests to
the STMs at up to approximately a 1kHz rate.  Samples in these
selection windows are then digitized and sent back to the DAQ crate.

Each STM has bi-directional trigger and data high-speed links to the
central DAQ which may utilize digital 1.2 Gbps or greater fiber optic
transceivers for noise immunity and speed.  The STM consists of 4
trigger/readout ASICs and an FPGA.  This FPGA handles the serial
transmission using the Xilinx Aurora protocol or similar.  Each of the
4 ASICs has 8 channels of input, and provides 8
Low-Voltage-Differential-Signal (LVDS) trigger outputs corresponding
to internal 1-shots that discriminate the full-RF-band trigger.
Carrier wave rejection in the trigger will be performed using
directional trigger masking.

Local pattern triggers among the feed-array signals are formed in the
FPGA, and broadcast via phase-offset encoding over the digital fiber
link.  Global triggers are formed in the DAQ master crate and
broadcast over the DAQ fiber uplink, resulting in a digitization
command.  The Trigger high-speed link is used for clock, timing and
commanding.

The digital data requested in response to a central DAQ module trigger
decision could arrive at the rate of up to 40~kHz, and must then be
further decimated since this rate is completely dominated by
incoherent thermal noise fluctuations rather than coherent plane-wave
impulses. A dedicated co-processor (which may just be a specific core
or pair of cores in a multi-core host processor) will make on-the-fly
reconstruction approximations to the digital waveform data and
determine which signals are consistent with coherent impulses to first
order. We expect to write data to the solid-state storage system at
no faster than about 50Hz, and this rate will still be dominated by
thermal noise, but will provide a continuous measure of instrument
health. Event size is estimated at about 40 kbyte per event
compressed, and this will yield about 9TB of data in a 50 day flight.

	\subsection{Metrology system.}

EVA's precursor mission ANITA had a mission critical
requirement for moderate accuracy orientation
measurements, to ensure that the free-rotation of the payload would
not preclude reconstruction of directions for events at the
degree level of accuracy. 
Such measurements were accomplished
with a redundant system of 4 sun-sensors, a magnetometer, and
a differential GPS attitude measurement system (Thales Navigation ADU5).
These systems performed well in flight and met the mission design
goals. In calibration done just prior to ANITA's 2006 launch, we
measured a total $(\Delta \phi)_{RMS} = 0.071^{\circ}$, very close to the limit of
the ADU5 sensor specification, and well within our allocated
error budget. 

For EVA, the issue of measuring the absolute orientation has shifted from the
payload position in the flight train to the balloon itself, along with the
inner feed membrane. To accomplish a measure of the orientation and
attitude of this system, we will use
a separate metrology system to locate fiducials
married to the surfaces of the outer balloon and inner membranes at strategic locations.
For orientation and attitude to be accurately measured from
the inner surfaces, including the reflector and feed array, we anticipate a
need for several-cm-level precision. We have demonstrated such precision already
using photogrammetry for the ANITA payload; implementation of this for EVA will require
periodic digital imaging of the balloon system from the BTP, with particular
attention to the fiducials. Laser or microwave ranging
systems may also have the capabilities necessary to augment or replace our photogrammetry
approach.

	\paragraph*{Timekeeping.}
Payload timekeeping for EVA will be required at several different levels. 
Dual GPS units will be used for absolute synchronization with UT
at the 50-100~ns level. One of these units also will be used to discipline
the onboard computer clock, for  millisecond-level
time tagging of events. A separate counter keeps track of 
nanosecond-level timing between digitizer boards and can link events
at longer time scales (of order 1 sec) as well. These systems are based on
successful systems used for both ANITA flights, and contain
various levels of redundancy which served ANITA well during both of its flights.

\section{Technical Challenges: Construction \& Launch.}

EVA presents challenging issues of construction and launching of
the balloon and feed membrane system. We have reviewed these issues
with NASA Columbia Scientific Balloon Facility engineers~\cite{CSBF1}.
The innovations associated with the EVA design all represent extensions
of technologies and methodologies that have one or more precursors in
existing balloon experience. For example, reflective radar tape is 
used for many balloons that are launched in regions where aircraft 
activity may be present to ensure that the radar cross section is
suitably large. Balloon-cap packages, while not common, are a relatively
accepted method of extending payload functionality when necessary.
Inner membranes have been utilized in the past for
a few discrete investigations, although none with the complexity of
the EVA membrane. However, this program will clearly require the
development of new production and assembly methods at the manufacturer
of the balloon (AeroStar International is the current NASA-sponsored
contractor for this), as well as new launch methods and associated hardware
at CSBF.

\section{EVA predictions for rates}

\subsection{UHE Neutrinos.}

\begin{table*}[hbt!]
\renewcommand{\baselinestretch}{1}
\begin{center}
\begin{small}
\caption{\small \it Expected numbers of events from a full range of BZ neutrino models
for a 50-day EVA flight compared to the combined sensitivity of ANITA-I and  ANITA-II,
which had a net 45 days of livetime. The last column gives the range of improvements in
the event totals.
\label{gzknus}}
\vspace{3mm}
  \begin{tabular}{|lccc|}
\hline 
{ {\it BZ neutrino models }}  &  {\it Events,}  & {\it Events,} & {\it ratio,}   \\   
                    &  {\it ANITA-II,28d} & {\it EVA,50d} & {\it EVA/ANITA}  \\ \hline
~~{Mixed UHECR composition}~\cite{Auger07a} & 0.05 &  5.0  & 100  \\
~~{Minimal, no evolution }~\cite{Engel01,PJ96,Aramo05} & 0.3-0.9 & 9.2-38 & $\sim 40$   \\
~~{$\Omega_m=0.3,\Omega_{\Lambda}=0.7$, Standard model}~\cite{Engel01} & 0.7 & 29 & 41 \\
~~{Waxman-Bahcall $E^{-2}$ flux (minimal)~\cite{WB}} & 0.49  & 6.5 & 13  \\
~~{GRB UHECR-sources}~\cite{Yuksel07} & 1.44 & 66 & 46 \\
~~{Strong source $z$-evolution}~\cite{Engel01,Aramo05,Kal02} & 2.2-5.3  & 40-60 & 11-18  \\
~~{Maximal, saturate all bounds}~\cite{Kal02,Aramo05}& 16-25 &  180-220 & $\sim 10$   \\ \hline
  \end{tabular}
 \end{small}
 \end{center}
 \renewcommand{\baselinestretch}{0.94}
\end{table*}

Using the results from our antenna simulation, and the basic
triggering scheme outlined above, we have developed a
Monte Carlo event simulation code to investigate whether,
using the frequency-dependent NEC2 model antenna gains described above,
EVA could achieve its minimum mission goal of achieving at
least one order of magnitude improvement in neutrino flux
sensitivity, with a desirable goal of two orders of magnitude
improvement. Since ANITA now has the best limits in our energy range,
we use the ANITA flux estimates as our baseline for comparison.
We estimated  effective neutrino apertures and event rates for a 50-day flight,
such as achieved by SPB flight 591NT.

Table~\ref{gzknus} presents the results of this simulation in terms of the
comparison of total neutrino events detected for a range of different models.
For these results, the energy threshold for EVA was improved by roughly
and order of magnitude compared to ANITA, giving the majority of triggered
neutrino events in the energy range from 0.3-3 EeV.
Our results indicate that EVA can improve current limits factors of 10-100
in total events detected, depending on the BZ neutrino 
model used. The lowest flux model, in the first line of the table, presents
a very difficult detection problem, and is based on a mixed-UHECR composition
as suggested by the Auger Observatory recent data~\cite{Auger07a}. EVA is able to
reach this challenging model, and all others that are current in the literature.
In the last line of the table, the models are in fact already
ruled out by ANITA results~\cite{ANITA2}, but we include this line for completeness.

\subsection{UHE Cosmic Rays.}

One of the most interesting results of the ANITA flights has been the serendipitous
observation of 16 or more ultra-high energy cosmic ray events, observed mostly
in reflection off the ice sheets, with their impulsive radio emission arising via the
geosynchrotron process in the Earth's magnetic field~\cite{ANITA_UHECR}. Due to
the largely vertical geomagnetic field in Antarctica, these events appear with
nearly pure horizontal polarization, and are thus cleanly separable from the
neutrino events, which are dominated by vertical polarization due to the Fresnel
geometry for refraction out from the ice to the balloon payload.

These events, whose mean energy approaches the GZK cutoff energy, signal a new methodology 
for probing the UHECR spectrum near the endpoint, which may be able to effectively
complement ground-based techniques. ANITA-III, due to fly in the 2013 season, and that will have a trigger
that is optimized for both UHE neutrino and UHECR detection, is
expected to observe several hundred such events, including of order 60 events above
the UHE energy at $10^{19}$~eV, and a handful at or above the
GZK cutoff energy at around $3 \times 10^{19}$~eV. 

Super-GZK events, those with energies
above $10^{20}$~eV, are a challenge for any ground-based observatory; in nearly
5 years' of observations, the Auger observatory has seen only of order 3 such events,
and ANITA-III, despite the significant increase over prior flights, still does not
effectively probe the super-GZK range. EVA, with much higher potential sensitivity than 
ANITA, may be expected to observe a much larger sample of UHECR events, and we have
simulated its potential sensitivity for UHECR detection, using simulations
developed for and validated by ANITA's UHECR analysis. We have also confirmed that ANITA
experimental measurements still do not agree well with any of the current ground-based
simulations, and thus we have used the experimental field strength estimates determined
by ANITA for our EVA simulation as well. Changes in the results are likely to appear
primarily in the energy scale rather than the total number of detected events, 
since that is the largest uncertainty in the ANITA field-strength estimators.

We find that EVA's total 
rate of detection of cosmic-ray-generated radio events will be a factor of 20
higher than that predicted for 
ANITA-III, with of order $15,000$ events detected in a 50-day flight,
a rate of 300 per day, compared to about 15 per day expected for ANITA-III~\cite{ANITA_UHECR}.
The vast majority of these events arise from relatively nearby lower-energy cosmic-rays,
with a mean energy of several EeV. Of order 500 of the sample have energies exceeding
$10^{19}$~eV, and of these, 60 or so are in the GZK cutoff range above $3 \times 10^{19}$~eV.
Only one or two events above $10^{20}$~eV may be expected due to the fact that,
despite the very large area -- of order $10^6$~km$^2$ -- observed by EVA, the small solid
angle for detection of the highly-beamed geosynchrotron radio emission leads to 
an effective saturation of the acceptance at high energies.
We expect to have initial energy resolution comparable to that estimated for
ANITA: $\Delta E/E \sim 50-100\%$. However, the much larger event sample will provide 
statistics in the region of the GZK cutoff (the only clear feature in an otherwise
smooth power-law distribution) that will enable the energy 
resolution to be calibrated and improved through analysis.

Such a large sample
of UHECR radio events, observed in the far-field, and with shower systematics and
geometries very different from those of ground-based observations,  
would represent a completely unique and independent measurement, and
provide a major boost in our understanding of the methodology of
UHECR detection via geosynchrotron radiation. 
This improved understanding could lead to development of
orbital platforms with far larger acceptance for super-GZK observations.

\section{Conclusions}

We find that a spherical-toroidal reflector system capable of observing Antarctica for the
detection of ultra-high energy particles with a very large aperture can be integrated into
a super-pressure balloon. Such a system could extend current sensitivity to UHE neutrinos
and cosmic rays up to two orders of magnitude, giving observatory-class performance for
these challenging particle astrophysics measurements. There are many challenges in the
construction, deployment, and data-processing for such a system, but these do not appear to
be fundamental limitations, and the science case for extending the reach of current methods
for UHE particle detection via such an approach appears to be compelling.

We thank NASA's Balloon Program Office and Columbia Scientific Balloon Facility,
and the Department of Energy's Office of Science for their support of these efforts.

 \setcounter{equation}{0}

\newcommand{\eone}{ {{\bf{e}}}_1(\phi) }
\newcommand{\etwo}{ {{\bf{e}}}_2(\phi) }
\newcommand{\ethree}{ {\bf{k}} }
\newcommand{\defmap}{ {\bf x}(s,\phi) }
\newcommand{\bfi}{ {\bf i}}
\newcommand{\bfj}{ {\bf j}}
\newcommand{\bfk}{ {\bf k}}
\newcommand{\mer}{{\bf n}_1(s,\phi)}
\newcommand{\cir}{{\bf n}_2(s,\phi)}
\newcommand{\ptder}[2]{\dfrac{\partial #1}{\partial #2}}


\clearpage
\appendix

\section{Pumpkin Geometry}

\def\PB{\mbox{{\sl Planetary Balloon}}}

      \begin{figure}
\centerline{\mbox{\psfig{figure=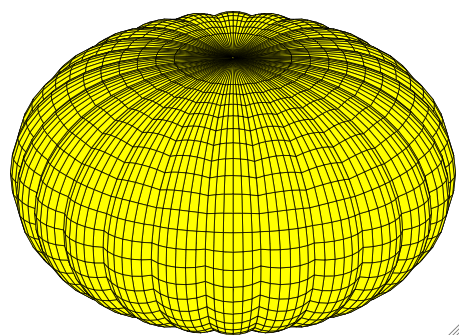,height=6.0cm}}} 
\caption{Complete pumpkin.}
\label{fig:completepump}
\end{figure}

The EVA  vehicle will be a super-pressure pumpkin balloon now 
in development by NASA's Balloon Program Office. See Fig.~\ref{fig:completepump} for an
illustration of a pumpkin balloon consisting of 24 gores.
The actual EVA pumpkin will likely be made of 
230 or more gores. The lobes will be shallower,
but still noticeable. In order to fully understand the geometry of such a structure
it will be beneficial to calculate curvatures for such a shape.

The design program {Planetary Ballon} \cite{RF}  developed by R. Farley, NASA/GSFC
will be used to determine the gore cutting pattern for the actual EVA balloon. 
In \PB, 
 mechanical properties of the balloon film and load tendons
 are used to estimate the strained  shape of a pumpkin lobe
 under certain design conditions.
 The flat cutting pattern for  the gore is then backed out from  this strained shape.
To analyze  the strained  EVA balloon, we will use the  analytical balloon model developed by
Baginski  (see, e.g.,  \cite{BBC}). 
 However, to simplify the exposition of the pumpkin 
geometry,  we will assume 
 the complete shape is cyclically symmetric and made of
$\nGores$ symmetric lobes and smooth within each lobe.   Within a typical  gore,   $\bfx(u,v)$ is a  parametrization  of a tubular
surface with generating curve $\Gamma$ defined by an Euler-Elastica curve (to be defined in the next section).  Proceeding in this
fashion, we obtain  a fairly
accurate representation of the equilibrium surface of a pressurized pumpkin balloon
constrained by
load tendons with a suspended payload \cite{BagSIAP05}.
The  tubular surface parametrization  will facilitate the discussion of balloon surface properties
that are relevant  to antenna applications.

\begin{figure}
\centerline{\mbox{\psfig{figure=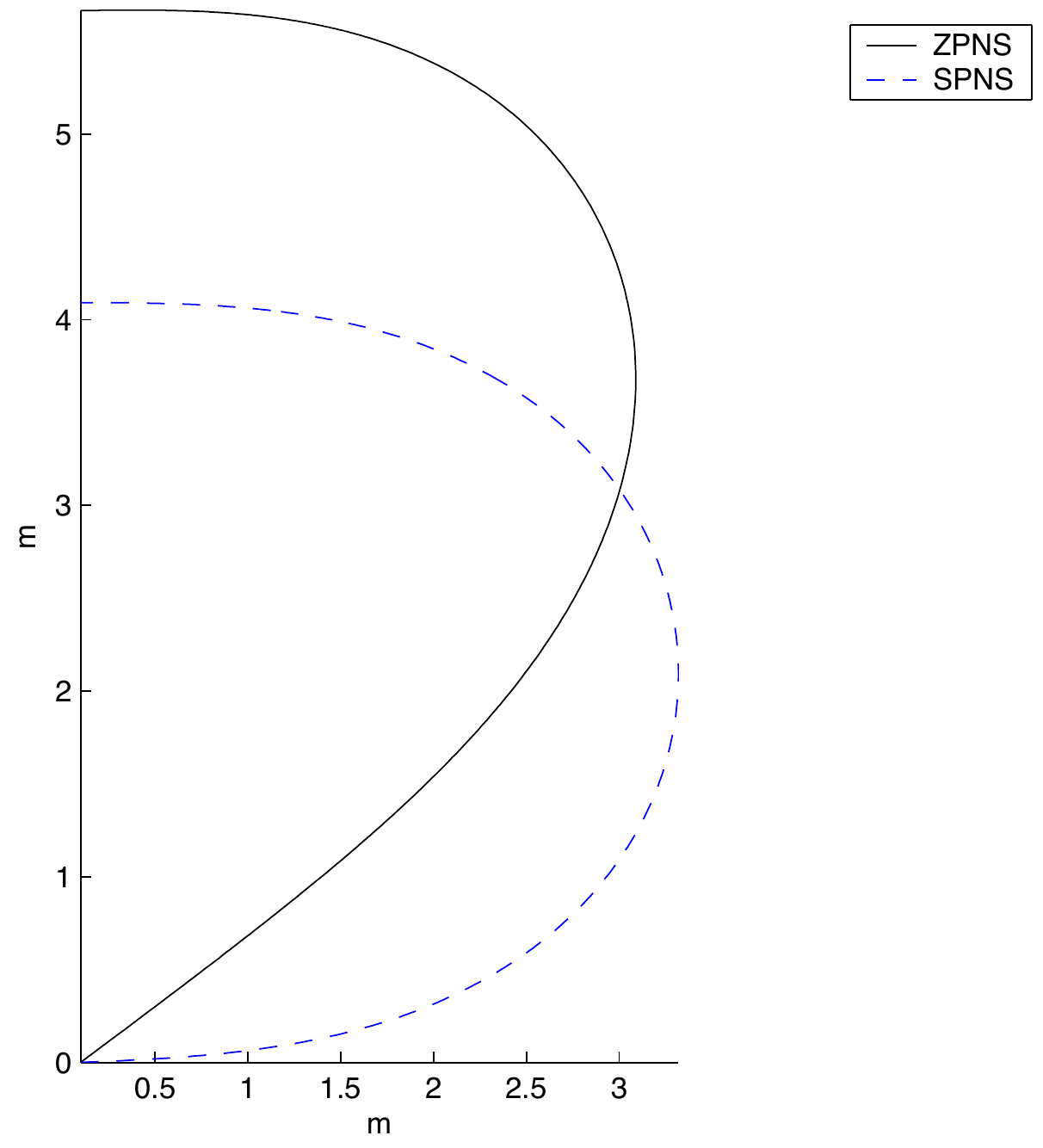,height=8cm}}} 
\caption{Comparison of a natural shape profile and the Euler-Elastica profile.}
\label{NSvsEE}
\end{figure}

\paragraph{Euler-Elastica}
The Euler-Elastica curve is the planar curve $(x, z) = (r(s), z(s))$ for  $0 < s < \ell$ satisfying \cite{BagSIAP05}
\be
\theta'' +  2 \tau \sin\theta =0,
\label{EE}
\ee
where 
$\theta$ is the angle between the tangent of the generating curve $\bft= (r'(s),z'(s)) = (\sin \theta, \cos\theta)$ and  $\bfk=(0,0,1)$. 
In this model,   we set the film weight density and tendon weight densities to zero  
(but include their total weight into the suspended payload) and set  $\tau = \pi p_0 /T_0 $ where 
 $T_0$ is the total tension (due to payload plus system weight)  and  $p_0$ is the constant  differential pressure.
Typical boundary conditions 
set  the   curvature  equal to zero  at the endpoints,
$
\theta'(0) = \theta'(\ell)=0$.
 The  modulus of the elliptic functions  associated with a
solution of Eq.~(\ref{EE})   is normally taken to be  $1/\sqrt{2}$.
In Fig.~\ref{NSvsEE}, we compare a natural shape profile (used in the design of zero-pressure balloons) with an Euler-elastica profile.
The radial component 
of the generating curve  for the elastica is
\be
r^2(\theta)= \tau^{-1}\sin(\onehalf\pi - \theta) 
\label{TAU}
\ee
 (see, \cite{Smc}) 
and the curvature of  the elastica curve is 
$\kappa(s) = \theta'(s) = -2 \tau r(s)$ (see \cite[Sec. 2.3.2]{BagSIAP05}). An
 expression for $z$ can be found by integrating  $z'(s)  = \cos\theta$
or by using elliptic functions \cite{MladenovOprea}.

\paragraph{Tubular Surface} 
We  present the  equations for a pumpkin shape balloon
as a  tubular surface.
We  assume that the  pumpkin shape  is made up of
$n_g$ symmetric pumpkin gores. 
 Let $\bfi=(1,0,0)$.
We begin with a curve,
$$\bfgamma(s)=r(s)\bfi  +  z(s) \bfk \in \realR^3,$$
 that 
we call the generator of the pumpkin gore.
We assume that $(r(s),z(s))$  is known from the solution of Eq.~(\ref{EE}).
$\bfgamma$
is 
parametrized by   arc length $s$,  i.e.,
$r'(s)^2 + z'(s)^2 = 1$.
Let $\bft$ denote the unit tangent of $\bfgamma$,   $\bfb$ its
inward unit normal
($\bfj = \bft \times \bfb$);
as defined in Eq.~(\ref{EE}), 
$\theta = \theta(s)$ is the angle  between $\bft$  and
$\bfk$,
and
\begin{eqnarray*}
\bft(s) &=& 
\sin\theta\bfi 
+
\cos\theta\bfj,\cr
\bfb(s) & = &
-\cos\theta\bfi
+
\sin\theta \bfj.
\end{eqnarray*}
The set $\lbrace \bfb, \bft, \bfj  \rbrace$ gives a right hand 
curvilinear basis for $\reals^3$.
Since $\bfgamma$ is a plane curve, its torsion is zero, and 
the Frenet equations  reduce to
\begin{eqnarray*}
\bft'(s) &=& \kappa(s) \bfb(s),\cr
\bfb'(s) &=& -\kappa(s) \bft(s)
\end{eqnarray*}
where $\kappa$ is the curvature of $\bfgamma$
(see \cite[Sec.~1.5]{doCarmo}).
We define 
a tubular surface in the following manner. For 
$-\pi< v < \pi, \ \ 0 < s < \ell$,  let
\be
\bfx( s, v) = \bfgamma(s) + \rBulge \left( -\bfb(s) \cos v + \bfj \sin v \right), \ \
\label{Tubular}
\ee
and
$x(s,v) =  \bfx(s,v) \cdot \bfi,
 y(s,v) = \bfx(s,v) \cdot \bfj$,  and $ z(s,v)=\bfx(x,v)\cdot \bfk$.

By direct calculation, we have
\begin{eqnarray*}
\bfx_s(s,v)  &= &(1 + \rBulge \kappa(s) \cos v) \bft(s), \\ 
\bfx_v(s,v) &=& \rBulge \left( \bfb(s) \sin v +  \bfj \cos v \right),\\
\bfx_s \times \bfx_v &=& \rBulge
(1+ \rBulge \kappa(s) \cos v) \cr
&&   \  \  \ \cdot  \left(     \bfb(s) \cos v  - \bfj \sin v  \right),
\end{eqnarray*}
and  area measure is $dS = \rBulge( 1 + \rBulge \kappa(s) \cos v) dsdv$.
A unit 
vector normal to the tubular 
surface is 
$$
\bfN(s,v) = \bfx_s \times \bfx_v  / | \bfx_s \times \bfx_v | =    \bfb(s)\cos v  - \bfj  \sin v.
$$
The triple $\{ \bfx_s, \bfx_v, \bfN  \}$
gives  a right hand basis for $\reals^3$.
Further calculation
yields, 
\begin{eqnarray*}
\bfN_s(s,v)  &=& -\kappa(s)  \bft(s) \cos v\cr
\bfN_v(s,v) & =& -\bfb(s) \sin v + \bfj \cos v.
\end{eqnarray*}
The principal curvatures are
\begin{eqnarray}
\kappa_1(s,v) & = & -\frac{\bfN_s \cdot \bfx_s}{\bfx_s \cdot \bfx_s } =
           \frac{\kappa \cos v}{1 + \rBulge  \kappa \cos v},  \label{eq:PrCurvaturesa}\\
\kappa_2(s,v)  &=& -\frac{\bfN_v \cdot \bfx_v}{\bfx_v \cdot \bfx_v} =
          \frac{1}{\rBulge}.\label{eq:PrCurvaturesb}
\end{eqnarray}
The tubular surface  is sufficiently smooth whenever
 (see \cite[p. 399]{doCarmo})
\be
\rBulge \kappa_0 <1, \mbox{ \  where  \ }
\kappa_0 < \max_{0 \le s \le \ell}|\kappa(s)|.
\label{REGULAR}
\ee
Condition
Eq.~(\ref{REGULAR}) is met in our applications.

\begin{figure}
\centerline{\mbox{\psfig{figure=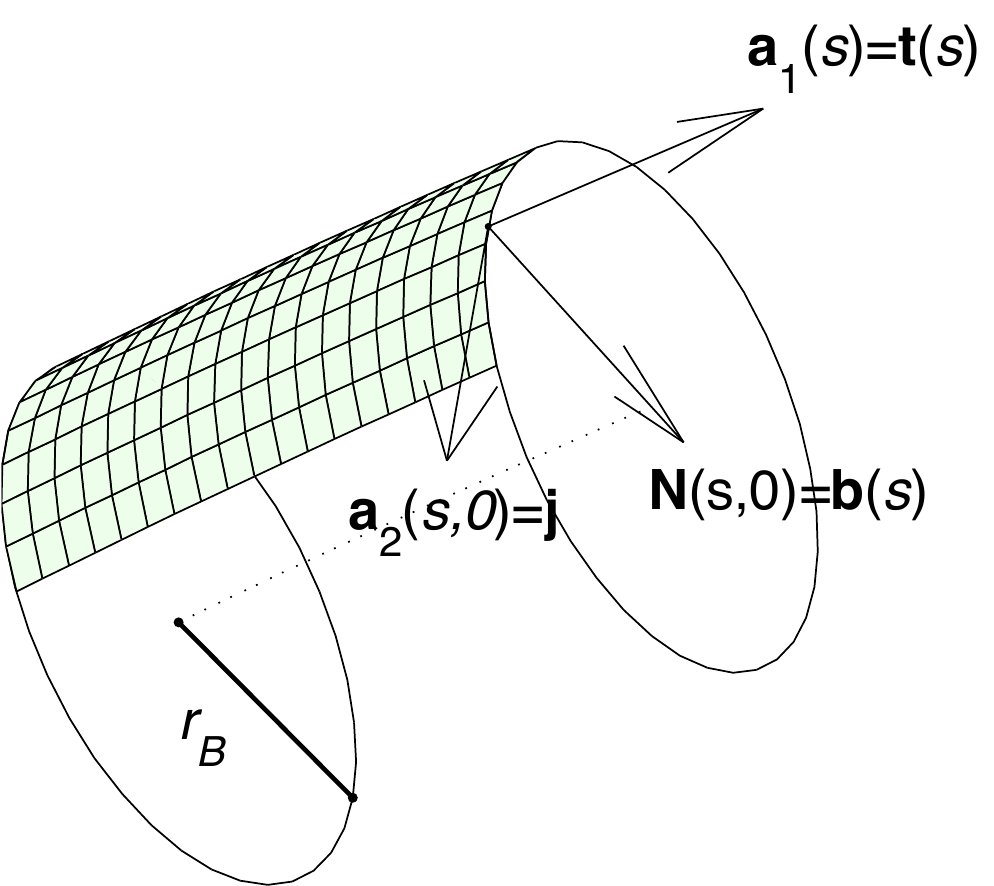,height=1.75in}}} 
\caption{  A pumpkin gore with $\lbrace \bfa_1, \bfa_2, \bfN\rbrace.$}
\label{GorePatch}
\end{figure}

A unit tangent to the curve  $s \rightarrow \bfx(s,v)$ is
$$
\bfa_1(s,v)  = \bfx_s(s,v)/|\bfx_s(x,v)|=\bft(s),
$$
and a unit tangent to the curve $v \rightarrow \bfx(s,v)$ is
$$ 
\bfa_2(s,v) =  \bfx_v(s,v)/|\bfx_v(s,v)| = \bfb(s) \sin v +\bfj \cos v.
$$
Note,
${\displaystyle
\frac{\partial \bfa_2}{\partial v} = \bfb(s) \cos v - \bfj \sin v = \bfN.}
$
In Figure~\ref{GorePatch},  the geometry of a pumpkin gore patch is illustrated
with the  vectors $\lbrace \bfa_1, \bfa_2, \bfN\rbrace$ when $v=0$.
Arc length in
the tubular surface along
a curve
parallel to the generator $s \rightarrow \bfx(s,v)$ is $\sbar$, where
$
d\sbar = (1 + \rBulge \kappa(s) \cos v) d s.
$

\begin{figure}
\centerline{\mbox{\psfig{figure=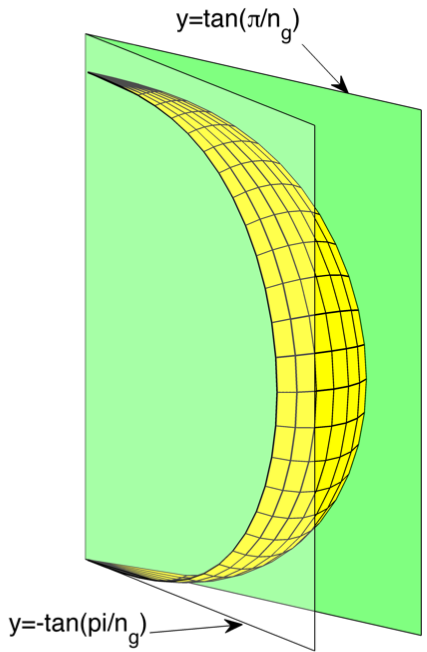,height=9.0cm}}} 
\caption{Pumpkin lobe.}
\label{fig:onegoretwoplanes}
\end{figure}

A pumpkin gore will be a subset of 
a tubular surface. We assume that 
the pumpkin gore is 
situated symmetrically with respect to the $xz$ plane
and interior to the wedge
defined by the half-planes $y = \pm \tan(\pi/\nGores) x$ with
$x \ge 0$. 
See Fig.~\ref{fig:onegoretwoplanes}.
We will refer to 
$\rBulge$ as the {\em  bulge radius} of the pumpkin gore.
The curve traced by  $ v  \rightarrow  
\bfgamma(s) +\rBulge( -\bfb(s) \cos v + \bfj \sin v)$
is a circle lying  in the plane with normal $\bft(s)$. To find the length
of the segment of the circle that forms a circumferential
arc of the pumpkin gore, 
we need to find the values of  $v$ where
this arc intersects the planes $y = \pm\tan(\pi/ \nGores) x$. 
For fixed $s$,
we find that $v$ must satisfy the 
condition 
$$
y(s,v) = \tan(\pi/n_g)~x(s,v). 
$$
This
leads us to the equation
\be
A(s) + B(s) \cos v + C \sin v =0,
\label{Pumpv}
\ee
where $C   =  \rBulge$, 
$B(s)   = -\rBulge \cos\theta(s) \tan(\pi/\nGores)$,
and
$A(s)  = - R(s)\tan(\pi/\nGores)$.
Eq.~(\ref{Pumpv}) can be solved for $v$, yielding
$$
v = v_g = \arccos\left(
\frac{ - A B +  C{\sqrt{C^2 + B^2 - A^2}}}{B^2 + C^2}
\right).
$$
Since $A$ and $B$ are functions of $s$ and other parameters,
so is 
$v_g =v_g(s, n_g, R(s), \theta(s))$.  
By symmetry,  the solution corresponding to the plane
 $y = -\tan(\pi/ \nGores) x$ is $v = - v_g$.
We define the 
three dimensional
pumpkin gore $\calG$ to be the  set,
\[
\calG = \left\lbrace  \bfx(s,v),  \ \  -v_g(s) < v < v_g(s), \ \
                   0 < s <  \ell
\right\rbrace.
\]
A complete shape $\calS$  has cyclic symmetry
and 
is made up of $\nGores$ copies of $\calG$.
In
 Fig.~\ref{fig:completepump}, we present a pumpkin
 balloon with $\nGores=24$, $\rBulge=10.9$~m,
 and $\max r = 55.77$~m. 
  Fig.~\ref{fig:completepump} was for demonstration purposes
 only.

\paragraph{Case Study}

\begin{figure}
\centerline{\mbox{\psfig{figure=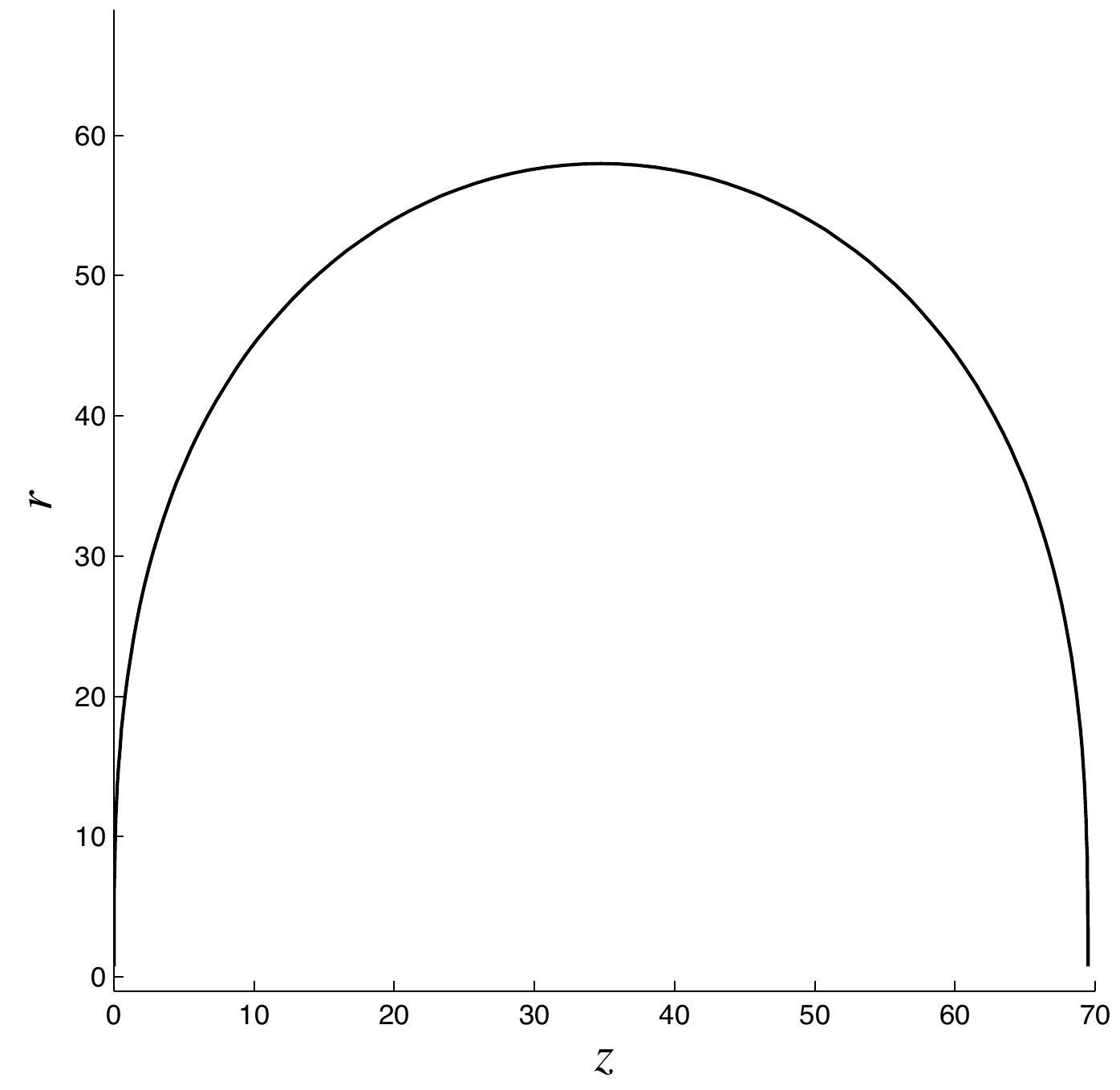,height=6.0cm}}} 
\caption{Euler-Elastica generator: $(z(s), r(s))$.}
\label{fig:gen}
\end{figure}

\begin{figure}
\centerline{\mbox{\psfig{figure=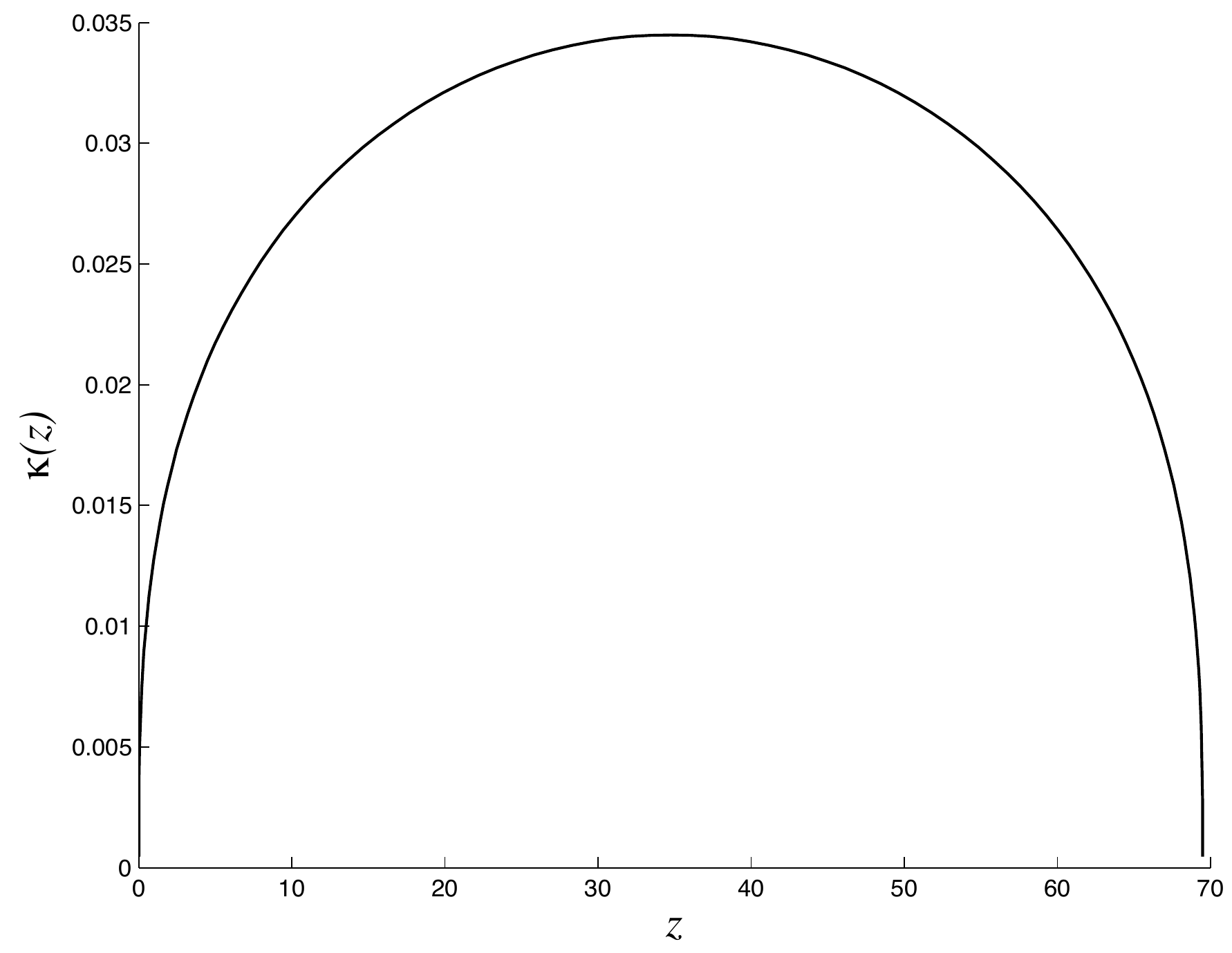,height=6.0cm}}} 
\caption{ $(z(s), \kappa(s))$ for $0<s<  \ell$.}
\label{fig:kappa}
\end{figure}

It will be beneficial to consider a pumpkin balloon
of sufficient size 
 for an EVA-type payload.  This leads to 
a balloon design that we can approximate as a tubular surface
as given by Eq.~(\ref{Tubular})
with $\nGores=230$ and  $\rBulge=2.7$~meters.
In this case we find the maximum radius of the
 generating curve   is  $\max r = 58$~meters and  $H= \max z = 69.49$~m.
In Figure~\ref{fig:gen}, we present the generating curve 
$(r(s), z(s))$ where $0< s <  \ell=302.6$~m for  the Euler-elastica  $\bfgamma$ as described. 
In Figure~\ref{fig:kappa}, we present   $(z(s), \kappa(s))$.
In Table~\ref{tab:equator}, we present a few data points located  with $\pm3.88$ meters of the equatorial plane.
$d(z,    z_0) $ measures the distance from the point $(r(z), z)$ to  the equatorial plane $z=z_0= 34.74$~m.
$\kappa(z)$ is the principle curvature at $z$. 

 If we rotate the
generating curve $\bfgamma$ about the  $z$ axis, the resulting surface will 
 have principle curvatures $\kappa$ and $\tilde\kappa$.
 The equatorial bulge angle is $2 v_b(z_{eq}) = 27.2$ deg.
This corresponds to an arc of $\bfgamma$ of approximately 11 meters in length.

A peculiarity of the  surface of revolution generated by 
an Euler-Elastica  is that its principal curvatures   $\kappa_m, \kappa_h$ 
satisfy  $\kappa_m(s,v) = \frac{1}{2}  \kappa_h(s,v)$. This is equivalent to the relation
   $R_h= 2 R_m$ where $\kappa_m = 1/R_m$ and $\kappa_h = 1/R_h$.
Such a shape is the basis for the popular mylar balloons and is discussed in
great detail in  \cite{MladenovOprea}. See Table~\ref{tab:equator}.
These observations will be of importance for the design of the 
 focusing reflector.

\begin{table}
\caption{Surface data in an equatorial band.    }
\begin{center}
\small
\begin{tabular}{ccccc}
\hline
 (m)& (m) &  (m)&       (m)& (deg)\cr
$d(z ,  z_{eq})$ & $r(z)$  &      $R_m$ &  $R_h$ & $v_b(z)$ \cr
\hline
 -3.8821& 57.7392   &   29.1310 & 58.2620 &    13.541 \cr
   -2.5924&   57.8839     &  29.0581&    58.1163&  13.575\cr
   -1.2975 &  57.9710    & 29.0145 &   58.0291&   13.596\cr 
         0  & 58.0000        &  29.0000  &   58.0000 &   13.603\cr
    1.2975&   57.9710  &   29.0145 &     58.0291&   13.596\cr
    2.5924 &  57.8839   &   29.0581 &   58.1163  &   13.575\cr
    3.8821 &  57.7392   &  29.1310 &     58.2620& 13.541\cr
    \hline
 \end{tabular}
\label{tab:equator}
\end{center}
\end{table}

\footnotesize
\vfil\eject

\end{document}